\documentclass[%
 aip,
 amsmath,amssymb,
 reprint,%
]{revtex4-1}
\usepackage{float}
\usepackage{graphicx}
\usepackage{dcolumn}
\usepackage{bm}

\usepackage[utf8]{inputenc}
\usepackage[T1]{fontenc}
\usepackage{mathptmx}
\begin{document}

\preprint{AIP/123-QED}

\title[]{Two-fluid model of rf current
condensation in magnetic
islands}

\author{S. Jin}
 \email{sjin@pppl.gov}

 \author{A. H. Reiman}%
 \email{areiman@pppl.gov}

\author{N. J. Fisch}
 \email{fisch@pppl.gov}
 
\affiliation{ Department of Astrophysical Sciences, Princeton University, Princeton, New Jersey 08543, USA}%
\affiliation{Princeton Plasma Physics Laboratory, Princeton, New Jersey 08540, USA}%

\date{\today}

\begin{abstract}
The stabilization of tearing modes with rf waves is subject to a nonlinear effect, termed rf current condensation, that has the potential to greatly enhance and localize current driven within magnetic islands. Here we extend previous investigations of this effect with a two fluid model that captures the balance of diffusive and thermal equilibration processes within the island. We show that the effective power, and resulting strength of the condensation effect, can be greatly enhanced by avoiding collisional heat loss to the ions. The relative impact of collisions on the overall power balance within the island depends on the ratio of the characteristic diffusion time scale and the electron-ion equilibration time, rather than the latter alone. Although relative heat loss to ions increases with island size, the heating efficiency does as well. In particular, we show that the latter safely dominates for large deposition profiles, as is typically the case for LHCD. This supports the possibility of passive stabilization of NTMs, without the precise aiming of the rf waves required for ECCD stabilization.

\end{abstract}

\maketitle

\section{Introduction}
Perturbations to the nested flux surfaces in tokamaks can result in magnetic islands, which greatly enhance radial transport and degrade confinement. In the absence of external heating, the pressure profile within the island tends to flatten. This locally suppresses the bootstrap current, reinforcing the original magnetic disturbance and  driving the island unstable. These unstable islands, known as neoclassical tearing modes (NTMs), are a major cause of disruptions\cite{de_Vries_2011, de_Vries_2014} and set a principal performance limit in tokamaks. \cite{ La_Haye_2006a,La_Haye_2009} 

Stabilization via current drive by rf waves\cite{Reiman_1983} has long been recognized as the leading solution, and has been the subject of much theoretical\cite{Hegna_1997,Zohm_1997,Yu_1998,Harvey_2001,Bernabei_1998,Kamendje_2005,La_Haye_2006b,La_Haye_2008,Sauter_2010,Smolyakov_2013,Ayten_2014,Borgogno_2014,F_vrier_2016,Li_2017,Grasso_2016} and experimental work.\cite{Zohm_1999,Prater_2003,Warrick_2000,Gantenbein_2000,Zohm_2001,Isayama_2000,La_Haye_2002,Petty_2004,Volpe_2015} Only recently, however, has the presence of the island and its effect on rf deposition been accounted for.\cite{Reiman_2018,Rodriguez_2019,eduardo,Jin_2020,nies_2020}

The thermal insulation provided by the closed magnetic topology and reduced cross field transport within the island\cite{Spakman_2008,Inagaki_2004,Bardoczi_2016,Ida_2012} produces significant temperature perturbations relative to the background plasma in the presence of rf heating. Relative temperature increases as large as 20\% have been reported.\cite{Westerhof_2007} Furthermore, the waves typically used, lower hybrid (LHW)\cite{fisch_1978} and electron cyclotron (ECW)\cite{fisch_1980}, resonate with superthermal electron populations at the location of the island and are thus highly temperature sensitive. The power that these waves can deposit is proportional to the number of resonant electrons ($P_{dep}\sim n_{res}$), which is exponentially sensitive to temperature perturbations ($n_{res}\sim \exp(-w^2)\approx \exp(-w_0^2)\exp(w_0^2\Delta T_e/T_{e0})$), where $w:=v_{res}^2/v_{th}^2$ is the ratio of the resonant and thermal velocities. As $w_0^2\approx4-20$ in practice\cite{kfj}, even small increases in temperature can drastically enhance the power absorption.

In combination, these properties create a positive feedback between the island temperature and rf deposition, resulting in significant nonlinear enhancement of current driven within the island. Since the temperature within the island is also governed by diffusion, the temperature will tend to be hottest in the center, thereby preferentially enhancing the driven current where it is most stabilizing. This amplification and focusing are termed the current condensation effect. \cite{Reiman_2018} The identification of this nonlinear process naturally leads to the suggestion that stabilization schemes should then be crafted to maximally exploit it. To this end, accounting for energy coupling between electrons and ions offers critical insights.

Previous investigations of the rf-condensation effect in magnetic islands have been confined to single-fluid models.\cite{Reiman_2018,Rodriguez_2019,eduardo,Jin_2020} Here we show that these are limiting cases of a more general two fluid picture---rf power injected in to the electrons may be lost in varying proportions through diffusive losses to the background plasma, or first through collisions with the ions. This energy balance is set not only by the strength of energy coupling between electrons and ions, but on the relative thermal diffusivities as well. 

Typical experimental parameters span a broad range of coupling strengths, and although the reduction of the electron thermal diffusivity within the island by an order of magnitude or more has been widely reported, similar results for the ions\cite{Ida_2012} are not as established. It is therefore of great value to establish an understanding of rf condensation that covers the wide variety of regimes relevant in practice.

We show here that the degree of energy coupling depends on the ratio of the characteristic electron diffusion time $t_{D.e}\sim W_i^2/\chi_e$ and electron-ion energy equilibration time $t_{eq}$, rather than the latter alone. The extent to which heating and stabilization are impacted by coupling depends on the ratio of diffusivities $\gamma:=\chi_e/\chi_i$; the effective power is reduced by a factor ($1+\gamma$) in the strongly coupled limit, compared to the decoupled limit. 

We also show that although both parasitic energy loss to the ions and effective power increase with the island width, the latter safely dominates. This suggests the possibility of self-healing islands for tokamaks with a significant fraction of the toroidal current sustained with lower hybrid waves (LHWs). We estimate the self-stabilization to occur at experimentally feasible rf power densities and island widths.

The paper is organized as followed. Section II introduces the two fluid model for the island temperatures, establishes limiting behaviors, and summarizes the essential features of the solutions. Section III characterizes the impact of energy coupling on the current condensation effect. Section IV discusses the possibility of self healing islands under LHCD. Section V summarizes the main results and conclusions.

\section{Two fluid model of island temperatures}
The energy transport equations for electrons and ions can be written as:

\begin{equation}
\label{eq:general}
    \frac{3}{2} \partial_t n_s T_s-\nabla \cdot ( n_s \chi_s \cdot \nabla T_s)=\frac{3}{2\tau_{eq}} n_s (T_{r}-T_{s})+P_{s}
\end{equation}
where subscript $s$ denotes either electrons or ions; subscript $r$ denotes the other species; $\chi_s$ is the heat diffusivity tensor of species $s$; and $\tau_{eq}=\frac{m_i}{m_e} \frac{3}{8}\sqrt{\frac{m_e}{2\pi}}\frac{(kT_e)^{3/2}}{n e^4 \lambda} $ is the electron ion equilibration time, where $\lambda\approx 20$ is the coulomb logarithm, $m_s$ is the mass of species $s$, $\mu$ is the ion mass in units of proton masses. $P_s$ contains whatever species specific heat sources and sinks may be present, aside from the explicitly written electron-ion equilibration term. 

 Considering rf induced temperature perturbations to an otherwise flat pressure profile within the island\cite{Maraschek_2012}, then $P_e=P_{rf}$ and $P_i=0$. Alternatively, even if the ohmic and radiation terms do not balance \cite{white_2015}, as long as they are negligible compared to the rf power, our selective choice of source terms is justified. As the time scales on which the island width or background plasma parameters evolve are typically much slower than the timescales of interest\cite{Rodriguez_2019} (i.e. the characteristic diffusion time of either species, $t_{D,s}:=3 W_i^2/8 \chi_{\perp,s}$, and the electron-ion equilibration time, $t_{eq}$), we may consider "steady state" solutions for which $\partial_t\rightarrow0$. On these timescales, the field lines will be approximately isothermal, so Eqs. (\ref{eq:general}) may be written as coupled 1-D diffusion equations for the perturbed temperatures $u_s:=w_0^2\widetilde{T_s}/T_{s,0}$:
\begin{equation}\label{eq:ue}
    \hat{D}u_e=P_{rf}+c(u_i-u_e)
\end{equation}
\begin{equation}\label{eq:ui}
    \gamma\hat{D}u_i=c(u_e-u_i)
\end{equation}
with $u_s=0$ at the island separatrix. Here $c:=t_{D,e}/t_{eq}$ is the ratio of the electron diffusion and electron-ion energy equilibration times, and $\gamma:= \chi_{\perp,i}/\chi_{\perp,e}$ is the ratio of the electron and ion diffusivities. The rf power term has been scaled to $P_{scl}:=n T_{e,0}/w_0^2t_{D,e}$. Further discussion of the boundary conditions and approximations used in arriving at Eqs. (\ref{eq:ue}) and (\ref{eq:ui}) can be found in Ref. \onlinecite{Rodriguez_2019}, with the difference here being the relaxed assumption of perfectly equilibrated ions and electrons.

 $\hat{D}$ is a diffusion operator that accounts for the geometry of the flux surfaces within the island\cite{Reiman_2018,allan}:
\begin{equation}
    \hat{D}:=-\frac{1}{\rho K(\rho)}\frac{d}{d\rho}\;\frac{E(\rho)-(1-\rho^2)K(\rho)}{\rho} \;\frac{d}{d\rho}
\end{equation}
where $\rho$ is a flux surface label that is 0 at the island center, and 1 at the separatrix. The island topology is more often defined with the alternate coordinate $\Omega:=8 (r-r_s)^2/W_i^2-\cos(m\xi)=2 \rho^2 -1$, where $r_s$ is the resonant radius, and $\xi:=\theta-\frac{n}{m}\phi$ is the helical phase, $\theta$ ($\phi$) and $m$ ($n$) are the poloidal (toroidal) angle and mode number respectively. \cite{De_Lazzari_2009,Bertelli_2011} 

The current condensation effect enters through the nonlinear rf heating term, $P_{rf}$. As discussed earlier, the exponential increase of superthermal resonant particles with small electron temperature leads to the following form: $P_{dep}\propto \exp(u_e)$. In the case where the rf deposition profile is wide compared to the island, the linear power deposition profile may be taken to be constant across the island region. This is especially appropriate for modeling lower hybrid current drive (LHCD) \cite{fisch_1987,kfj}. For such a power bath model, the rf power term takes the form:
\begin{equation}
    P_{rf}=P_0\exp(u_e)
\end{equation}
This model isolates the physics of RF-condensation, but neglects the possibility of ray depletion.

The two-fluid equations \ref{eq:ue} and \ref{eq:ui} capture the flow of energy through the island system---in through rf heating of the electrons, and out through diffusive losses from both the electrons and ions. The total energy loss from the electron population is the sum of two competing processes, direct diffusive losses to the environment, and collisional heat exchange with the ions. The relative strength of each process is roughly characterized by the dimensionless parameter $c:=\tau_{D,e}/\tau_{eq}\sim W_i^2$. Of course, heat loss to the ions is also regulated by how quickly the ions can dump this heat to the environment---this is characterized by the parameter $\gamma:=\chi_i/\chi_e$.

\subsection{Limiting cases}

The two fluid model of the island temperature contains 3 characteristic time scales: the electron diffusion time $\tau_{D,e}$, the ion diffusion time $\tau_{D,i}$, and the electron-ion energy equilibration time $\tau_{eq}$. The ratios of the first quantity to the latter two give the two dimensionless coupling parameters: $\gamma=:\chi_{\perp,i}/\chi_{\perp,e}=\tau_{D,e}/\tau_{D,i}$ describing the relative diffusivities of the ions to the electrons, and $c:=\tau_{D,e}/\tau_{eq}$ describing the relative strength of collisional to diffusive processes for the electrons. Familiarity with the limiting cases of these coupling parameters will build a foundation for understanding further two-fluid results, and contextualize the single fluid models that have been used thus far. 
\subsubsection{Single fluid reduction in limiting collisionality regimes}
If there is significant scale separation of the electron diffusion time and equlibration time, the dynamics are essentially that of a single fluid.
In the $c\rightarrow0$ limit, only the electron dynamics are relevant, and the collisional term, along with the ion equation can be dropped entirely.

The $c\rightarrow\infty$ limit corresponds to the electrons and ions having identical temperatures and behaving as a single fluid with an averaged conductivity. This can be seen by writing equation \ref{eq:ui} in a more illuminating form:
\begin{equation}
    u_i=u_e-\frac{1}{c}(\gamma \hat{D} u_i)
\end{equation}

so as $c \rightarrow \infty$, we can take $u_i \approx u_e=u$.

Adding the ion and electron equations then gives
\begin{equation}\label{eq:single}
    \frac{1+\gamma}{2}\hat{D} u= P_{rf}/2
\end{equation}
so the uncoupled electron equation is recovered with a new diffusivity that is the average of the electron and ion diffusivities, and a halved source term. The exact form of equation \ref{eq:ue} with $c=0$ can be obtained by substituting new scalings ($t_{scl} \rightarrow t_{scl} 2/(1+\gamma) ,  \quad P_{scl} \rightarrow P_{scl} (1+\gamma)$. We see that in this limit, the system behaves as a single fluid with the power effectively reduced by the factor $1/(1+\gamma)$. Technical exceptions to these limits will be discussed in the following section.

\subsubsection{Influence of conductivity ratio $\gamma$}
The above single fluid cases can be further refined depending on the relative ordering of the ion-diffusion time. 
The $c\rightarrow0$ limit will generally correspond to completely cold ions, with the electron dynamics unaffected;i.e. for the orderings (1)$\tau_{D,e}<<\tau_{D,i}<<\tau_{eq}$ ($c\Rightarrow0,\;\gamma\rightarrow0$) and (2)$\tau_{D,i}<<\tau_{D,e}<<\tau_{eq}$ ($c\rightarrow0,\;\gamma\rightarrow\infty$). When both parameters are going to the same limit, the double arrow denotes the one that approaches that limit faster. 

For the special case of (3)$\tau_{D,e}<<\tau_{eq}<<\tau_{D,i}$ ($c\rightarrow0,\;\gamma\Rightarrow0$) , although $c\rightarrow0$, the ions will eventually also arrive at the same temperature as the electrons. This heating occurs despite the collisional heat source to the ions vanishing, as the rate at which they diffusively dissipate any input heat is vanishing even faster. This is a particularly pathological case however, and only mentioned for completeness.

For $c\rightarrow\infty$, the ions will generally be at the same temperature as the electrons. This holds true for the orderings (4)$\tau_{eq}<<\tau_{D,e}<<\tau_{D,i}$ ($c\rightarrow\infty,\;\gamma\rightarrow0$) and (5)$\tau_{eq}<<\tau_{D,i}<<\tau_{D,e}$ ($c\Rightarrow\infty,\;\gamma\rightarrow\infty$), but case (6)$\tau_{D,i}<<\tau_{eq}<<\tau_{D,e}$ ($c\rightarrow\infty,\;\gamma\Rightarrow\infty$) is subtly different. Case (4) is the only one of the $c\rightarrow\infty$ cases that can accomplish efficient heating, since $\gamma\rightarrow0$ the ions piggyback on the electron temperature for free. 

The steady state result of case (4) is identical to case (3) of the $c\rightarrow0$ set, but in contrast, case (4) has the electrons and ions at the same temperature on timescales short compared to the evolution of the of the electrons, while case (3) would require times scales long compared to the electron diffusive timescale for the ions to catch up. 

Cases (5) and (6) both correspond to both species being ineffectively heated, thanks to the scaling of power by the factor ($1+\gamma$). The subtle difference between the two cases is that the relative temperature difference in case (6) can be significant, as although the directly heated electrons are immediately exchanging their heat with the ions, the ions are dumping that heat even faster. 

\subsection{Steady state solutions}
\begin{figure}[h]
\centering     
\includegraphics[width=\linewidth]{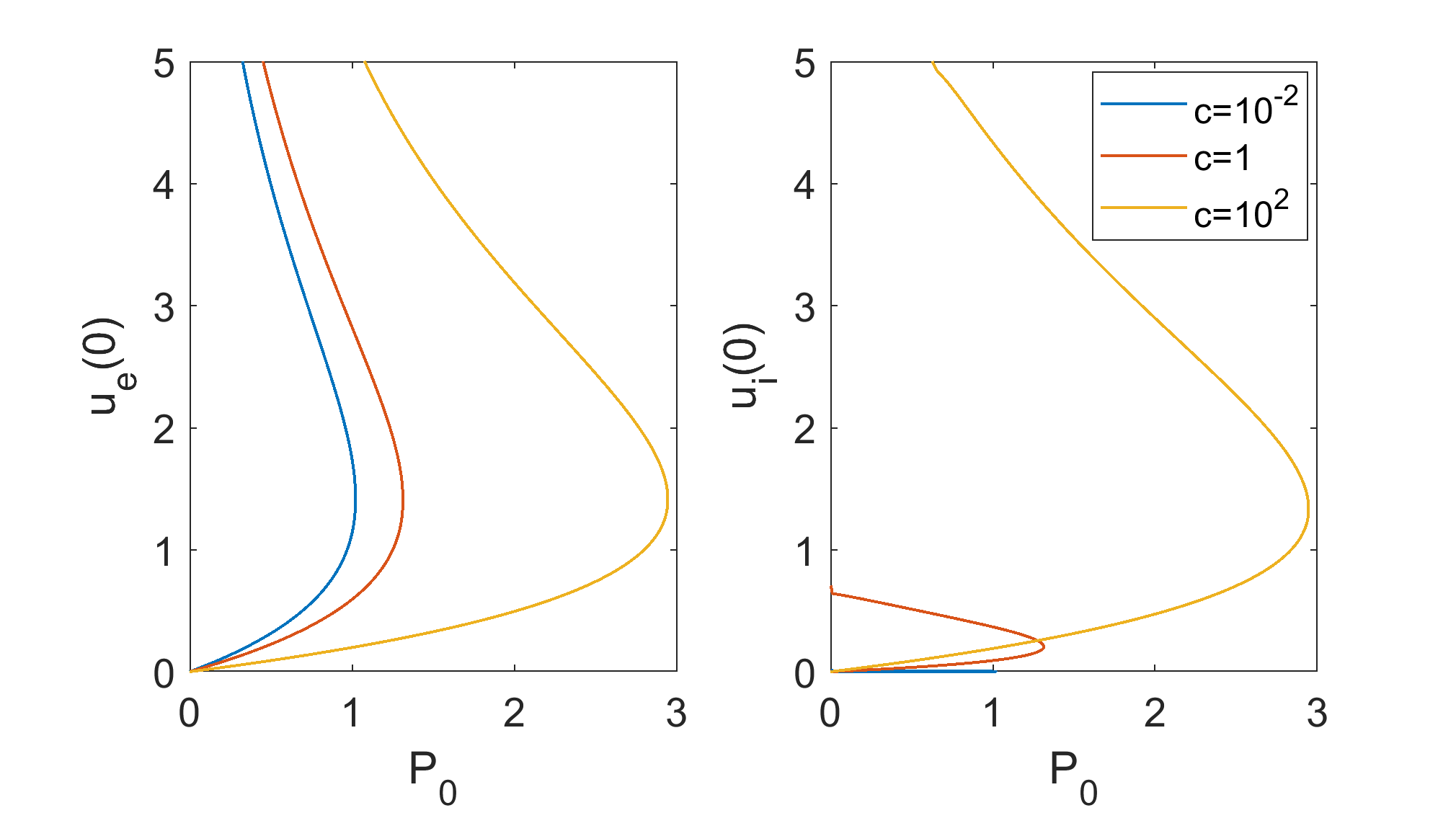}
\caption{\label{fig:us}Electron and ion temperatures vs scaled power $P_0$, for $\gamma=2$.}
\end{figure}

Prior to discussing the impact of two fluid features in particular, here we will highlight the basic properties of rf current condensation, common to all coupling regimes. As there is typically a large separation between the time scales of interest \cite{Rodriguez_2019}, i.e. diffusion or energy equilibration, and the MHD timescales on which the island evolves, we can discuss "steady state" solutions for which the island width can be treated as a constant. Eqs. (\ref{eq:ue}) and (\ref{eq:ui}) then yield two solution branches, joined at a bifurcation point, as shown in Fig. \ref{fig:us}. The lower (upper) branch is stable (unstable),

As discussed in more detail in Ref. \onlinecite{Reiman_2018}, the power bath model does not admit steady state solutions for rf powers past the bifurcation point. At these higher powers, the island temperature will continue to grow until encountering additional physics not included here, such as ray depletion\cite{Rodriguez_2019} or stiffness. \cite{eduardo} Note that power localization improves with increasing island temperature, further enhancing the stabilization efficiency. The bifurcation point therefore provides an estimate of the rf power required to access dramatic nonlinear enhancement, for a given set of plasma/island parameters. The sudden temperature increase (relative to the MHD time scales on which the island typically evolves) also serves to provide a distinctive experimental signature. 
 
It must be noted however, that the bifurcation point is by no means a requirement for stabilization, as will be discussed further in section IV. Since the power deposition is exponentially enhanced by the temperature perturbation, sufficient stabilization can occur at moderate electron temperatures, especially with the higher efficiency (larger $w_0^2$) of LHWs. This could be beneficial for avoiding turbulent transport enhancement at larger island temperatures\cite{eduardo}.

\section{Impact of energy coupling on stabilization}
\begin{figure}[h]
\centering     
\includegraphics[width=\linewidth]{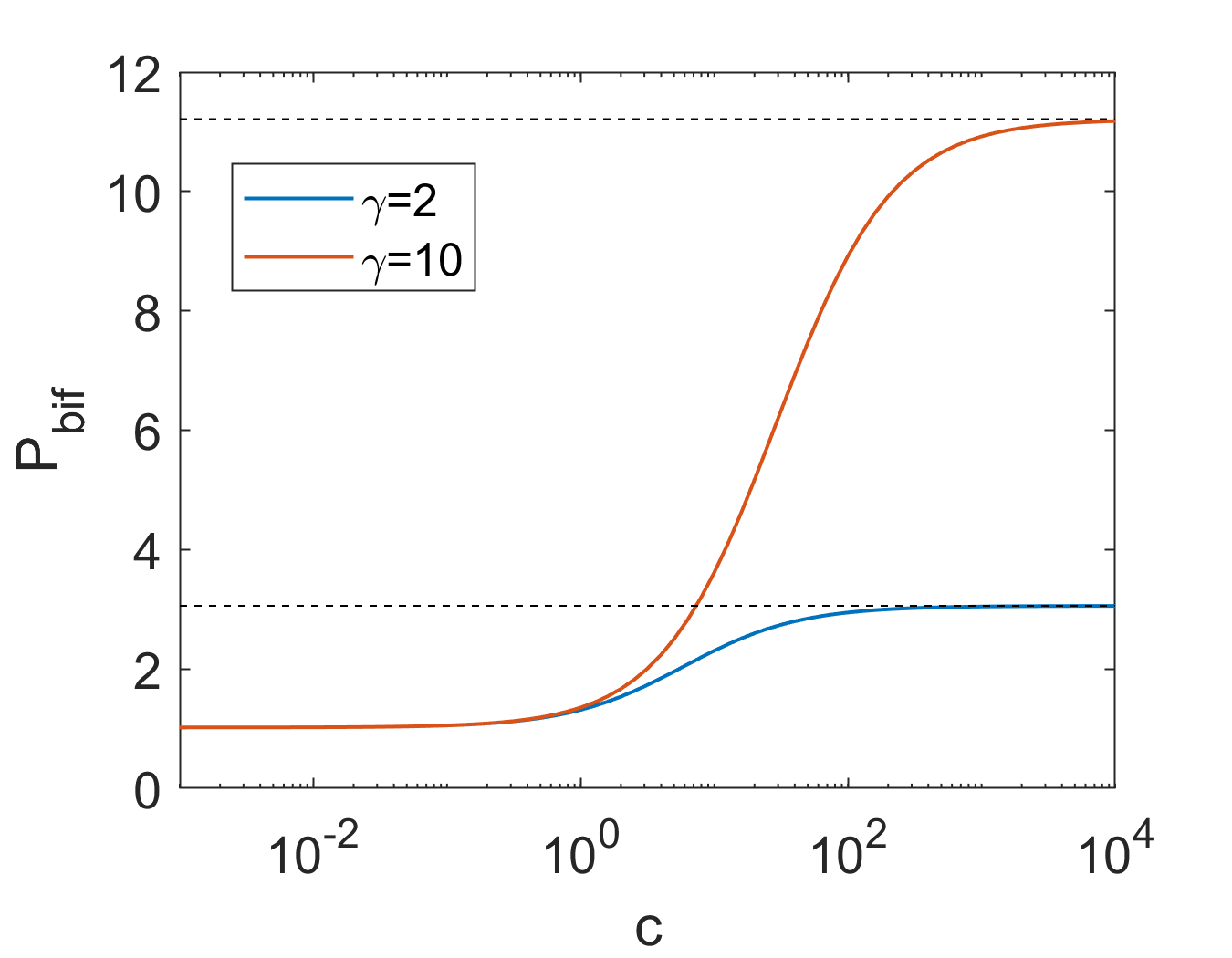}
\caption{\label{fig:pbif} Scaled power to reach bifurcation point vs. coupling parameter $c$.}
\end{figure}

As shown in our discussion of limiting cases, the effective power is reduced by a factor of $1+\gamma$ when the electrons are fully equilibrated with the ions, relative to when their energies are decoupled. At minimum, $\gamma\approx2$ in the case of turbulent transport\cite{Erba_1998}, and could be much higher, as turbulence suppression and reduced electron diffusivities have been widely reported in islands\cite{Spakman_2008,Inagaki_2004,Bardoczi_2016}. In particular, if transport is predominantly neoclassical, then $\gamma\approx10$.  As the power is nonlinearly related to the island temperature, this can amount to an even more significant impact on stabilization. Fig. \ref{fig:pbif} summarizes the effect of energy coupling on the effective power, using the bifurcation threshold as a marker.

It can be seen that the dimensionless power corresponding to the bifurcation threshold scales roughly logarithmically with coupling parameter $c$ between the limiting regimes, $c \approx10^{-2} - 10^2$. Although Fig. \ref{fig:pbif} provides the most concise summary of the impact of energy coupling on effective power, it must be emphasized that $P_{bif}$ and $c$ are not the actual power and equilibration rate, but are scaled to $P_{scl}=nT_{e0}/w_0^2 t_{De}$ and $t_{De}^{-1}$ respectively. To get a sense of the typical values for magnetic islands in present day devices, with $n\approx10^{13}-\;3\times 10^{13}\; cm^{-3}$, $T_{e0}\approx 0.5\;-\;3 \;keV$, $W_i\approx 5-30 \;cm$, $\chi_e\approx 0.1\;-\; 1\;m^2/s$, this gives $P_{scl}\approx0.1-10 \;KW/m^3$ and $c\approx 0.1-10$. Lower effective power, both through collisions with the ions, and through the increase of $P_{scl}$ can therefore be a concern for higher density plasmas. Although higher $T_{e0}$ and smaller $W_i$ reduce the relative heat loss to ions, the increase of $P_{scl}$ will dominate, leading to lower effective power.

The stabilization efficiency also depends on the current profile, with current driven closer to the center being more stabilizing, while current driven closer to the periphery can even be destabilizing. Fortunately, the shape of the power deposition profile is notably insensitive to the central heat sink provided by the ions. It will require a larger effective power to achieve a certain $u_e(0)$, but for that $u_e(0)$ the rest of the temperature (and resulting power) profile will not be appreciably different. Note that this would not be the case for narrow, off center profiles which could result from imprecisely aimed ECCD. Energy coupling can significantly impact the temperature profiles in such cases (Appendix C), exacerbating the risk of misalignment, on top of the reduced effective power. It should therefore be appreciated that there is one less thing to worry about for the broad current profiles considered here, as the reduced effective power is decisively the dominant effect of energy coupling on stabilization. 

\section{Self-healing under LHCD}
It has long been understood that localized current is essential for stabilization\cite{Reiman_1983}. Only recently, however, has it been shown that such localized deposition can be accomplished even with initially broad profiles, via the nonlinear enhancement of deposition with electron temperature. Traditional (linear) calculations do not capture this natural focusing. As a result, ECCD has received the vast majority of attention for NTM stabilization, while LHCD has been largely ruled out due to its broader deposition profiles. This must be reconsidered in light of the current condensation effect. 

Although ECCD is valued for its ability to deposit power in a steerable region smaller than the island width\cite{Zohm_2007}, the stabilization enhancement of this sharp localization is contingent upon precise aiming of the rf. This poses a significant challenge in practice as it requires active control techniques.\cite{Volpe_2009, Nelson_2019} In stark contrast, rf condensation opens the possibility of fully passive stabilization in steady state tokamaks where a significant fraction of the toroidal current is driven with LHWs. 

Significant theoretical progress has been made on this front. Simulations performed with the GENRAY ray tracing code coupled to the CQL3D Fokker Planck solver demonstrated significant localization of LHCD for typical electron temperature profiles within the island.\cite{Frank_2020} A new code, OCCAMI, coupling GENRAY to a thermal diffusion equation solver within the island to provide higher fidelity simulations of the current condensation effect, is in development.\cite{nies_2020} This tool has already been used to show the feasibility of accessing the bifurcation point and related hysteresis phenomena in an ITER-like scenario with realistic parameters. 

The focused treatment of energy coupling presented here allows us to further expand the presently single-fluid based understanding of rf condensation. As will be shown shortly, energy coupling is a critical feature to take into account, especially for the investigation of self-healing under LHCD. We have shown in this case that the increase of effective power with island width ($P_0 \sim W_i^2$) dominates the increasing energy losses to the ions (which scales at most logarithmically with $W_i^2$). As a result, the stabilizing current naturally increases as the island heats up and grows. 

At some island width, this may be enough to entirely stabilize the island. The width at which this occurs may be estimated as follows. The modified Rutherford equation (MRE) gives the island width growth rate as a function of different driving and stabilizing mechanisms\cite{De_Lazzari_2009,Bertelli_2011}:
\begin{equation}
    \frac{\text{d}w}{\text{d}t}\propto \Delta_0'(w)-\Sigma_i\Delta_i'(\delta j_i)
\end{equation}
where $\Delta_0'$ is the classical stability index, and $\Delta_i'(\delta j_i)$ are corrections to the classical tearing mode equation due to perturbations of the parallel current at the resonant surface. These corrections have the form \cite{De_Lazzari_2009}:

\begin{equation}
\begin{aligned}
        \Delta_i & \propto \int_{-\infty}^{\infty}\text{d}x\oint\text{d}\xi \cos(\xi)\delta j_i (x, \xi)\\
         & =\int_{-1}^{1}\text{d}\Omega\int_{-\hat{\xi}}^{\hat{\xi}}\text{d}\xi \frac{\cos(\xi)}{\sqrt{\Omega+\cos(\xi)}}\delta j_i (\Omega, \xi)
\end{aligned}
\end{equation}
where $\hat{\xi}=\cos^{-1}(-\Omega)$, $x:=r-r_s$ is the radial displacement from the resonant surface, $\Omega$ is the flux surface label and $\xi$ is the helical phase, as defined in Section II.

The dominant driving term for NTMs will be due to the perturbed bootstrap current $\delta j_{bs}$\cite{Westerhof_2016}, while the stabilizing term of interest for our purposes comes from the perturbed LH current $\delta j_{LH}$. Then we estimate stabilization to occur when $0\approx \Delta'_{bs}+\Delta'_{LH}$. 

We take the perturbed bootstrap current to be constant over the island, $\delta_{bs}=-J_{bs}$, where $J_{bs}$ is the bootstrap current density at the resonant surface prior to the island formation:
\begin{equation}
    \Delta'_{bs}\propto -J_{bs}\int_{-1}^{1}\text{d}\Omega F(\Omega)
\end{equation}
 where $F(\Omega):=\int_{-\hat{\xi}}^{\hat{\xi}}\text{d}\xi \frac{\cos(\xi)}{\sqrt{\Omega+\cos(\xi)}}$ serves as a flux surface weighting function.

The perturbed rf current $\delta_{LH}$ is given by the nonlinear enhancement to the LH current at the resonant surface, due to the the heating of the island. The current driven is roughly proportional to the power deposition, neglecting the dependence of the current drive efficiency on temperature, which is negligible compared to the exponential enhancement factor\cite{fisch_1987}:
\begin{equation}
    \Delta'_{LH}\propto J_{LH}\int_{-1}^1\text{d}\Omega (\exp(u_e(\Omega))-1)F(\Omega)
\end{equation}
where $J_{LH}$ is the LH current at the resonant surface prior to the island formation. As before, the flux surface label $\Omega$ is related to the coordinate $\rho$ in Eqs. (\ref{eq:ue}) and (\ref{eq:ui}) by $\rho^2=(\Omega+1)/2$.

The stabilization condition can finally be written by balancing the contributions of the LH and bootstrap currents:
\begin{equation}
    1=R^1\frac{\int_{-1}^1\text{d}\Omega (\exp(u_e(\Omega))-1)F(\Omega)}{\int_{-1}^{1}\text{d}\Omega F(\Omega)}
\end{equation}
where $R:=J_{bs}/J_{LH}$ is the ratio of the bootstrap to LH currents prior to island formation.

\begin{figure}[h]
\centering     
\includegraphics[width=\linewidth]{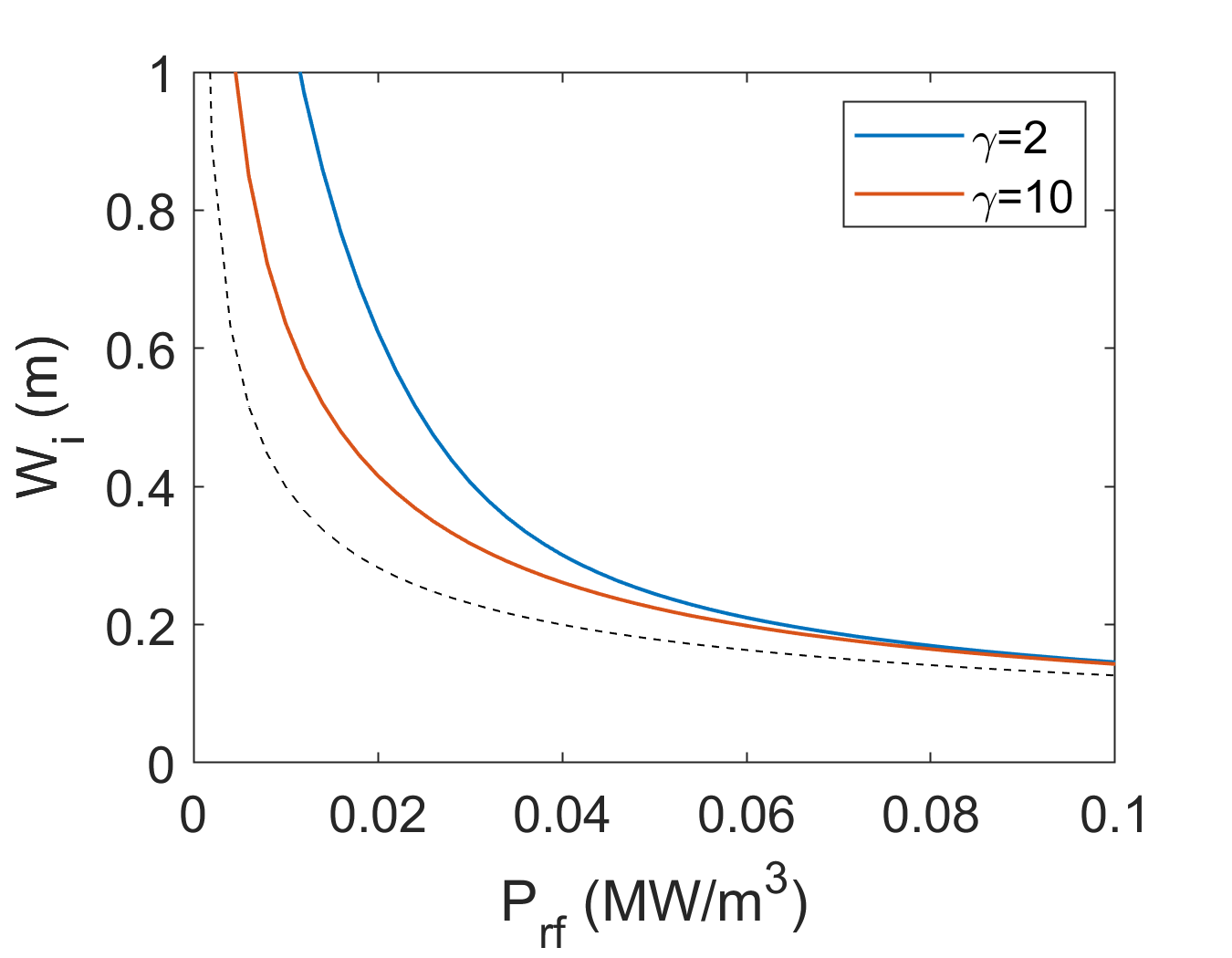}
\caption{\label{fig:wstab} Island width $W_i$ at self-stabilization vs. rf power density $P_{rf}$, calculated for $R=1$. Dotted black line indicates $c\rightarrow0$ limit.   Calculated for $w_0^2=10$, $\chi_e=0.1 \;m^2/s$, $n=10^{20} \;m^{-3}$, $T_{e0}=4 \;keV$.}
\end{figure}

As an example, Fig. \ref{fig:wstab} shows the island width at which the balance between the LH and bootstrap driving terms is reached as a function of ambient (pre-island) rf power density, with plasma parameters anticipated for ITER Scenario 2 at the q=2 surface, as studied in Ref. \onlinecite{Frank_2020}. Unsurprisingly, higher powers will allow islands to be stabilized at smaller widths, by counteracting the decreased confinement within the island. Of course, Fig. \ref{fig:wstab} can just as easily also be interpreted as the amount of rf power required to reach the threshold for an island of a certain size. It should be noted that the island sizes and rf power densities corresponding to the threshold are in an experimentally relevant range\cite{Frank_2020}.

It can be seen that at lower powers, due to the large island widths required to reach the threshold, there can be significant impact from energy losses to ions. In this regime, there is also a strong dependence on the ratio of diffusivities $\gamma$. Due to the suppression of turbulent transport within the island, it may well be the case that $\gamma$ is closer to its neoclassical value At higher powers, the stabilization width may be reached at smaller island widths, and the effects of energy coupling are negligible. As typical experimental parameters span both strong and weak coupling regimes, a two fluid analysis is indispensable. 

Note that the island width at stabilization will take the form $W_i\sim C/\sqrt{P_rf}$ in the high or low power limits, with the constant $C$ differing by a factor of $\sqrt{1+\gamma}$. Consequently, there will be diminishing gains from additional power input, in terms of the island width at stabilization. Physically, this is a result of smaller islands requiring steeper edge gradients, and thus more power input, to support a given central temperature that is needed for significant nonlinear enhancement.
 \begin{figure}[h]
\centering     
\includegraphics[width=\linewidth]{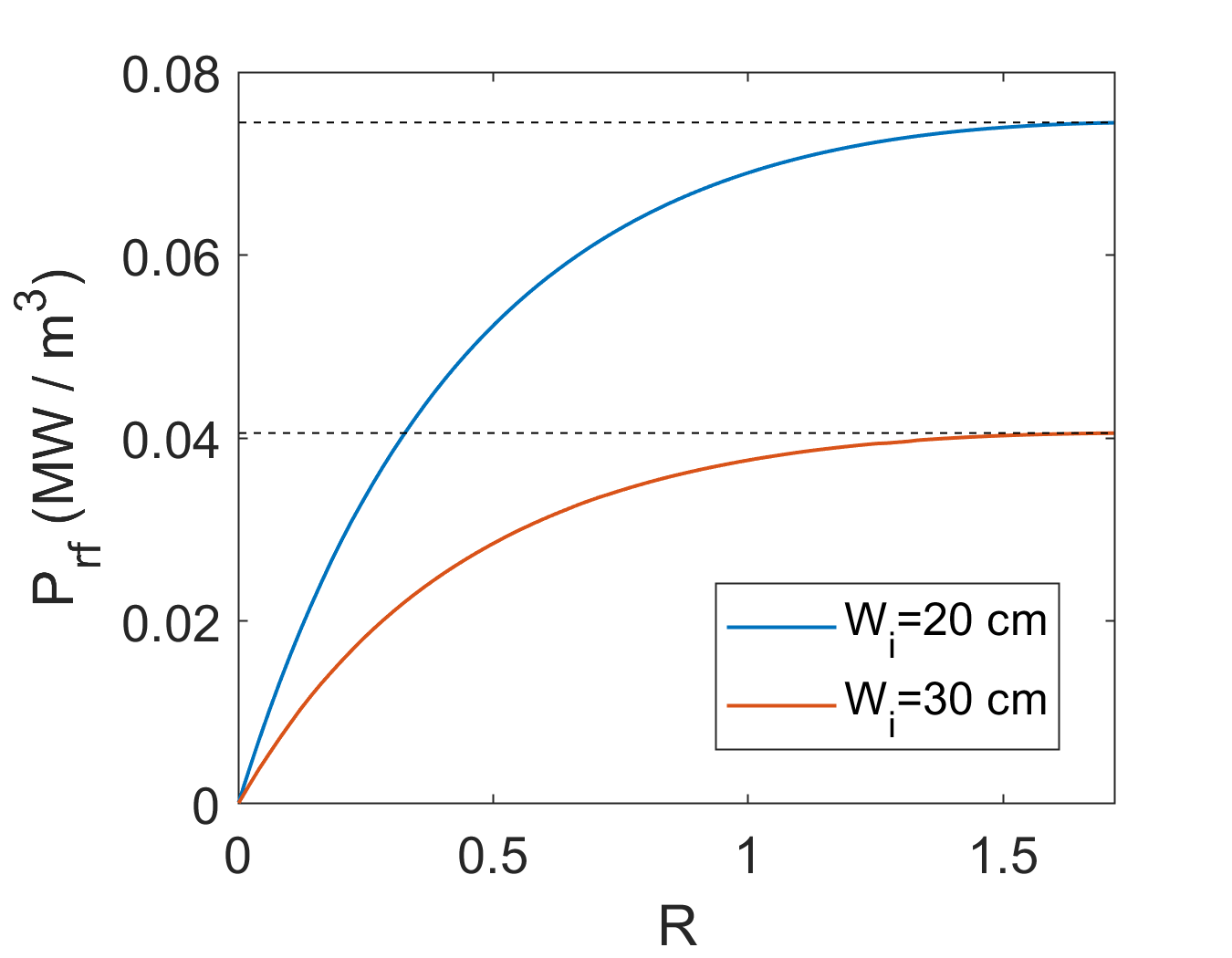}
\caption{\label{fig:pstab} RF power density required for stabilization vs. bootstrap-LH fraction R. Dotted black line indicates bifurcation power for each island width. Calculated for $w_0^2=10$, $\chi_e=0.1 \;m^2/s$, $n=10^{20} \;m^{-3}$, $T_{e0}=4 \;keV$,$\gamma=2$.}
\end{figure}

As can be seen in Fig. \ref{fig:pstab}, the power required for stabilization at a target width is only appreciably sensitive to the bootstrap-LH current ratio for $R\lessapprox1$, with the power requirement barely increasing as the bifurcation power is approached. This is simply due to the rapid growth of the island temperature close to the bifurcation point. For $R\gtrapprox1.5$, stabilization requires island temperatures past the bifurcation point. If the island grows past the bifurcation point, the island temperature and resulting power deposition will rapidly increase until reaching a hotter solution branch. A hysteresis effect may then be accessed---once on this upper branch, the rf stabilization will be greatly enhanced via the current condensation effect, and the island may then be suppressed to a smaller saturated width than would otherwise be possible without reaching the bifurcation point.

Therefore, a smaller bootstrap to LH ratio could be desirable in the sense that the island will be stabilized at smaller widths than the width at bifurcation. On the other hand, a larger bootstrap to LH ratio would allow the hysteresis effect to be accessed. Especially considering the insensitivity of the stabilization island width to higher powers, and weak dependence of the requisite stabilization power to the bootstrap-LH current ratio, an interesting optimization issue is raised. If smaller saturated island widths are made possible by passing the bifurcation point and jumping to the hotter branch of solutions, it could potentially be more efficient to use a lower power such that the island is able to grow to the threshold width and then experience enhanced absorption, rather than a higher power that stabilizes the island below the threshold but at a larger final size. While this clearly depends on just how much better the upper branch performs (which would depend on saturation mechanisms not examined in this work), even more moving parts are introduced by the inter-dependencies between $P_0$, $w_0^2$, and $R$. Altogether, this poses a fascinating problem that calls for additional modeling and experimental investigations.

\section{Summary}
The impact of energy coupling between electrons and ions on the current condensation effect has been reported, using a two temperature diffusive model. In the limit of perfectly equilibrated electrons and ions, the effective power is reduced by a factor of $(1+\gamma)$ where $\gamma:=\chi_i/\chi_e$ is the ratio of the ion to electron thermal diffusivities, relative to the case where the electron and ion temperatures are entirely decoupled. The strength of energy coupling is characterized by the dimensionless quantity $c:=t_{D,e}/t_{eq}$ where $t_{D,e}=3 W_i^2/8\chi_e$ and $t_{eq}$ are the characteristic electron diffusion time and electron-ion equilibration time, respectively. Although energy coupling to the ions increases with the island width, it is found that the resulting reduction in effective power is overwhelmed by the improved heat confinement in larger islands. Thus, in the presence of broad LHCD profiles, stabilization may be achieved passively as an island grows, with higher rf powers triggering the condensation effect at smaller island widths. The popular dismissal of LHWs for instability control must therefore be reconsidered.

\begin{acknowledgements}

This work was supported by Nos. U.S. DOE DE-AC02-
09CH11466 and DE-SC0016072.
\end{acknowledgements}
\section*{Data Availability}
The data that support the findings of this study are available from the corresponding author
upon reasonable request.

\appendix

\section{Impact of energy coupling for narrow deposition}
Here we examine the impact of energy coupling for the case of narrow deposition profiles that are fully contained within the island, as would be typical of ECCD. Since there is no additional power absorbed as the island grows, it can be immediately anticipated that the relative impact of increased energy coupling will be greater in this case, than for broad deposition profiles. A narrow deposition profile also introduces the possibility of misalignment, which would further offset the improved energy confinement provided by a larger island. We explore the impact of these effects with Eqs. (\ref{eq:ue}) \& (\ref{eq:ui}), but in slab geometry $\hat{D}\rightarrow \partial_x^2$ where $x=2(r-r_s)/W_i$, and with a highly simplified deposition profiles:
\begin{equation}\label{eq:source}
    P_{rf}=P_{0}\delta(x)
\end{equation}
The nonlinear factor of $\exp(u_e)$ does not enter here, as the power is already fully deposited and maximally localized. The source term Eq. (\ref{eq:source}) admits the following solutions:
\begin{equation}
    u_e=\frac{P_0}{2(1+\gamma)}[1-|x|+\frac{\gamma}{k}\frac{\sinh(k(1-|x|)}{\cosh(k)}]
\end{equation}
\begin{equation}
    u_i=\frac{P_0}{2(1+\gamma)}[1-|x|-\frac{1}{k}\frac{\sinh(k(1-|x|))}{\cosh(k)}]
\end{equation}
where $k:=\sqrt{c(1+\gamma^{-1}}$, and $P_0$ is the linear power density scaled to $P_{scl}=n T_{e0} W_i / 2 \tau_{De}$. Note the additional factor of $W_i/2$ in the power scaling compared to the broad deposition case.

 \begin{figure}[h]
\centering     
\includegraphics[width=\linewidth]{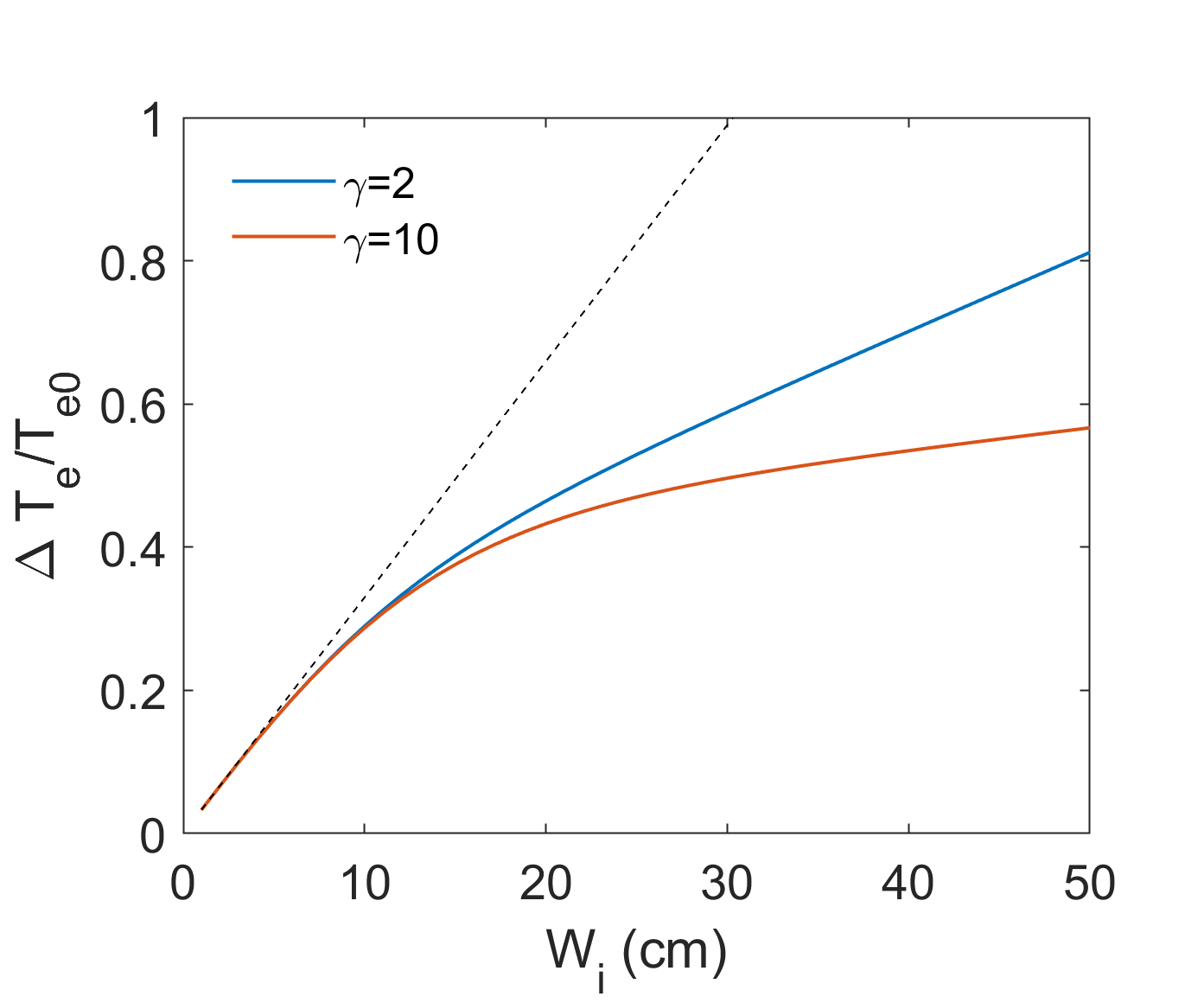}
\caption{\label{fig:eccalc}Central island temperature  vs. island width for 20 MW of rf power deposited at the $q=2$ surface in ITER ($n=10^{20}\; m^{-3}, T_{e0}=5 \;keV, \chi_e=0.1\; m^2/s, r=1.5\; m$). Dashed line indicates single-fluid solution neglecting collisions. $c\approx 1$ at $W_i=10 \;cm$.}
\end{figure}

It can be immediately seen that both single fluid limits, $c\rightarrow0$ and  $c\rightarrow\infty$ reduce to $u_e(0)\rightarrow P_0/2$ and $u_e\rightarrow P_0/2(1+\gamma)$ respectively. It should be noted however, that although $W_i\rightarrow\infty$ does correspond to $c\rightarrow\infty$, the limiting solution in this case does not simply have the effective power reduced by the $1+\gamma$ factor as in the broad deposition case, but there is an additional offset as $P_0\sim W_i$ and the second term of the solutions $\sim W_i^{-1}$. This can be seen in Fig. \ref{fig:eccalc}, which shows how the central electron temperature $u_e(0)$ scales with island width for ITER scenario II parameters.

\section{Narrow off-center deposition in intermediate coupling regimes}
Energy coupling has a uniquely detrimental effect for narrow, off-center power deposition in the intermediate coupling (two fluid) regime. Steady state solutions of the single fluid equations satisfy an equation of the form $-u''=P$, where $P>=0$ throughout the island. This, along with the symmetry enforced by $x$  and $-x$ corresponding to the same flux surface, means that the maximum point in the temperature profile $u(x)$ will be at $x=0$ (center of the island). This central peaking constraint can be broken in intermediate collisionality regimes, when the temperature difference $c(u_i-u_e)$ exceeds the power deposited at the center. 

 Considering two $u_e$ profiles of equal area (so same total amount of heating accomplished), the one with the central dip will have larger edge gradients. Accordingly, this dipped profile will experience larger diffusive losses and cost more input power to maintain. A centrally dipped profile also means that the power deposition at the periphery will be enhanced at the expense of the center. Thus we may anticipate that intermediate collisionality regimes can perform poorly beyond what may be anticipated from simply sharing power with the ions. 
 
 This can be simply demonstrated with the following toy profiles:
 \begin{equation}\label{eq:source}
    P_{rf}=P_{0}(\delta(x-\Delta)+\delta(x+\delta))/2
\end{equation}
where $\Delta\in[0,1)$ represents the degree of misalignment. The symmetrized form is to account for the points at $\pm x$ belonging to the same flux surface. Again, the nonlinear factor of $\exp(u_e)$ does not enter here, as the power is already fully deposited and maximally localized. The source term Eq. (\ref{eq:source}) admits the following solutions:
\begin{equation}
    u_e=\frac{P_0(1-\Delta)}{2(1+\gamma)}
       \begin{cases}
            \frac{1+x}{1-\Delta}+\gamma a \frac{\sinh(k (1+x))}{\sinh(k (1-\Delta))} &\text{$x<\Delta$}\\
     1+\gamma a \frac{\cosh(k x)}{\cosh(k \Delta)} &\text{ $|x|<\Delta$} \\
     \frac{1-x}{1-\Delta}+\gamma a \frac{\sinh(k (1-x))}{\sinh(k (1-\Delta))} &\text{$x>\Delta$}
   \end{cases}
\end{equation}
\begin{equation}
    u_i=\frac{P_0(1-\Delta)}{2(1+\gamma)}
       \begin{cases}
            \frac{1+x}{1-\Delta}- a \frac{\sinh(k (1+x))}{\sinh(k (1-\Delta))} &\text{$x<\Delta$}\\
     1- a \frac{\cosh(k x)}{\cosh(k \Delta)} &\text{ $|x|\leq\Delta$} \\
     \frac{1-x}{1-\Delta}- a \frac{\sinh(k (1-x))}{\sinh(k (1-\Delta))} &\text{$x>\Delta$}
   \end{cases}
\end{equation}
where $k:=\sqrt{c(1+\gamma^{-1}}$, $a:=[k(1-\Delta)(\tanh(k\Delta)+\coth(k(1-\Delta))]^{-1}$, and $P_0$ is the linear power density scaled to $P_{scl}=n T_{e0} W_i / 2 \tau_{De}$. 

 \begin{figure}[h]
\centering     
\includegraphics[width=\linewidth]{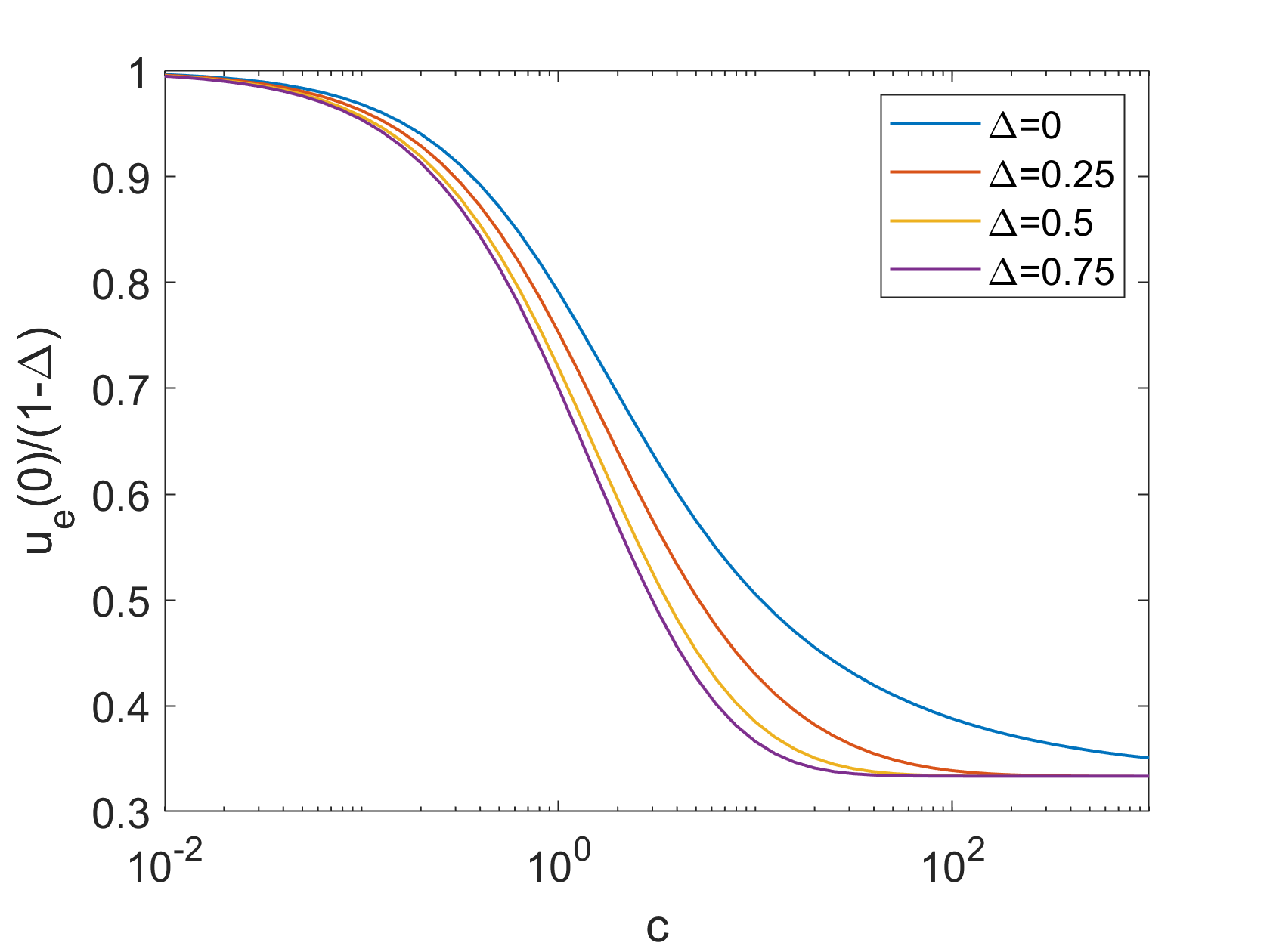}
\caption{\label{fig:os}Central island temperature (normalized for uncoupled impact of misalignment) vs. coupling parameter c for various $\Delta$. }
\end{figure}
In the absence of coupling ($k\rightarrow0$), the misalignment simply scales down the central electron temperature by a factor of $1-\Delta$. Fig. \ref{fig:os} shows the impact of coupling on the central island temperature for various degrees of misalignment, after normalizing for the impact of misalignment in the uncoupled limit by dividing out a factor of $1-\Delta$. It can be seen that higher $\Delta$ exacerbates the power reduction due to coupling for intermediate coupling strengths ($c\sim 1-10$), while eventually reducing its negative influence to the linear factor of $1-\Delta$ in the single fluid limits.

It should be emphasized that this hyper-simplified calculation is intended to illustrate the potential for uniquely inefficient heating for off-center profiles in the intermediate coupling regime, but not to quantify how large this effect may be. For a nonlinear calculation incorporating a self consistent treatment of ray damping and island heating, this intermediate coupling effect may be expected to be larger than what is seen here, as the deposition can be pulled further into the periphery of the island as it heats up (see discussion of shadowing in \onlinecite{Rodriguez_2019} or \onlinecite{Jin_2020}),

\nocite{*}

\bibliography{aipsamp}

\providecommand{\noopsort}[1]{}\providecommand{\singleletter}[1]{#1}%
\begin{thebibliography}{55}%
\makeatletter
\providecommand \@ifxundefined [1]{%
 \@ifx{#1\undefined}
}%
\providecommand \@ifnum [1]{%
 \ifnum #1\expandafter \@firstoftwo
 \else \expandafter \@secondoftwo
 \fi
}%
\providecommand \@ifx [1]{%
 \ifx #1\expandafter \@firstoftwo
 \else \expandafter \@secondoftwo
 \fi
}%
\providecommand \natexlab [1]{#1}%
\providecommand \enquote  [1]{``#1''}%
\providecommand \bibnamefont  [1]{#1}%
\providecommand \bibfnamefont [1]{#1}%
\providecommand \citenamefont [1]{#1}%
\providecommand \href@noop [0]{\@secondoftwo}%
\providecommand \href [0]{\begingroup \@sanitize@url \@href}%
\providecommand \@href[1]{\@@startlink{#1}\@@href}%
\providecommand \@@href[1]{\endgroup#1\@@endlink}%
\providecommand \@sanitize@url [0]{\catcode `\\12\catcode `\$12\catcode
  `\&12\catcode `\#12\catcode `\^12\catcode `\_12\catcode `\%12\relax}%
\providecommand \@@startlink[1]{}%
\providecommand \@@endlink[0]{}%
\providecommand \url  [0]{\begingroup\@sanitize@url \@url }%
\providecommand \@url [1]{\endgroup\@href {#1}{\urlprefix }}%
\providecommand \urlprefix  [0]{URL }%
\providecommand \Eprint [0]{\href }%
\providecommand \doibase [0]{http://dx.doi.org/}%
\providecommand \selectlanguage [0]{\@gobble}%
\providecommand \bibinfo  [0]{\@secondoftwo}%
\providecommand \bibfield  [0]{\@secondoftwo}%
\providecommand \translation [1]{[#1]}%
\providecommand \BibitemOpen [0]{}%
\providecommand \bibitemStop [0]{}%
\providecommand \bibitemNoStop [0]{.\EOS\space}%
\providecommand \EOS [0]{\spacefactor3000\relax}%
\providecommand \BibitemShut  [1]{\csname bibitem#1\endcsname}%
\let\auto@bib@innerbib\@empty
\bibitem [{\citenamefont {de~Vries}\ \emph {et~al.}(2011)\citenamefont
  {de~Vries}, \citenamefont {Johnson}, \citenamefont {Alper}, \citenamefont
  {Buratti}, \citenamefont {Hender}, \citenamefont {Koslowski},\ and\
  \citenamefont {and}}]{de_Vries_2011}%
  \BibitemOpen
  \bibfield  {author} {\bibinfo {author} {\bibfnamefont {P.}~\bibnamefont
  {de~Vries}}, \bibinfo {author} {\bibfnamefont {M.}~\bibnamefont {Johnson}},
  \bibinfo {author} {\bibfnamefont {B.}~\bibnamefont {Alper}}, \bibinfo
  {author} {\bibfnamefont {P.}~\bibnamefont {Buratti}}, \bibinfo {author}
  {\bibfnamefont {T.}~\bibnamefont {Hender}}, \bibinfo {author} {\bibfnamefont
  {H.}~\bibnamefont {Koslowski}}, \ and\ \bibinfo {author} {\bibfnamefont
  {V.~R.}\ \bibnamefont {and}},\ }\bibfield  {title} {\enquote {\bibinfo
  {title} {Survey of disruption causes at {JET}},}\ }\href {\doibase
  10.1088/0029-5515/51/5/053018} {\bibfield  {journal} {\bibinfo  {journal}
  {Nuclear Fusion}\ }\textbf {\bibinfo {volume} {51}},\ \bibinfo {pages}
  {053018} (\bibinfo {year} {2011})}\BibitemShut {NoStop}%
\bibitem [{\citenamefont {de~Vries}\ \emph {et~al.}(2014)\citenamefont
  {de~Vries}, \citenamefont {Baruzzo}, \citenamefont {Hogeweij}, \citenamefont
  {Jachmich}, \citenamefont {Joffrin}, \citenamefont {Lomas}, \citenamefont
  {Matthews}, \citenamefont {Murari}, \citenamefont {Nunes}, \citenamefont
  {Pütterich}, \citenamefont {Reux},\ and\ \citenamefont
  {Vega}}]{de_Vries_2014}%
  \BibitemOpen
  \bibfield  {author} {\bibinfo {author} {\bibfnamefont {P.~C.}\ \bibnamefont
  {de~Vries}}, \bibinfo {author} {\bibfnamefont {M.}~\bibnamefont {Baruzzo}},
  \bibinfo {author} {\bibfnamefont {G.~M.~D.}\ \bibnamefont {Hogeweij}},
  \bibinfo {author} {\bibfnamefont {S.}~\bibnamefont {Jachmich}}, \bibinfo
  {author} {\bibfnamefont {E.}~\bibnamefont {Joffrin}}, \bibinfo {author}
  {\bibfnamefont {P.~J.}\ \bibnamefont {Lomas}}, \bibinfo {author}
  {\bibfnamefont {G.~F.}\ \bibnamefont {Matthews}}, \bibinfo {author}
  {\bibfnamefont {A.}~\bibnamefont {Murari}}, \bibinfo {author} {\bibfnamefont
  {I.}~\bibnamefont {Nunes}}, \bibinfo {author} {\bibfnamefont
  {T.}~\bibnamefont {Pütterich}}, \bibinfo {author} {\bibfnamefont
  {C.}~\bibnamefont {Reux}}, \ and\ \bibinfo {author} {\bibfnamefont
  {J.}~\bibnamefont {Vega}},\ }\bibfield  {title} {\enquote {\bibinfo {title}
  {The influence of an iter-like wall on disruptions at {JET}},}\ }\href
  {\doibase 10.1063/1.4872017} {\bibfield  {journal} {\bibinfo  {journal}
  {Physics of Plasmas}\ }\textbf {\bibinfo {volume} {21}},\ \bibinfo {pages}
  {056101} (\bibinfo {year} {2014})}\BibitemShut {NoStop}%
\bibitem [{\citenamefont {Haye}\ \emph {et~al.}(2006)\citenamefont {Haye},
  \citenamefont {Prater}, \citenamefont {Buttery}, \citenamefont {Hayashi},
  \citenamefont {Isayama}, \citenamefont {Maraschek}, \citenamefont {Urso},\
  and\ \citenamefont {Zohm}}]{La_Haye_2006a}%
  \BibitemOpen
  \bibfield  {author} {\bibinfo {author} {\bibfnamefont {R.~L.}\ \bibnamefont
  {Haye}}, \bibinfo {author} {\bibfnamefont {R.}~\bibnamefont {Prater}},
  \bibinfo {author} {\bibfnamefont {R.}~\bibnamefont {Buttery}}, \bibinfo
  {author} {\bibfnamefont {N.}~\bibnamefont {Hayashi}}, \bibinfo {author}
  {\bibfnamefont {A.}~\bibnamefont {Isayama}}, \bibinfo {author} {\bibfnamefont
  {M.}~\bibnamefont {Maraschek}}, \bibinfo {author} {\bibfnamefont
  {L.}~\bibnamefont {Urso}}, \ and\ \bibinfo {author} {\bibfnamefont
  {H.}~\bibnamefont {Zohm}},\ }\bibfield  {title} {\enquote {\bibinfo {title}
  {Cross{\textendash}machine benchmarking for {ITER} of neoclassical tearing
  mode stabilization by electron cyclotron current drive},}\ }\href {\doibase
  10.1088/0029-5515/46/4/006} {\bibfield  {journal} {\bibinfo  {journal}
  {Nuclear Fusion}\ }\textbf {\bibinfo {volume} {46}},\ \bibinfo {pages}
  {451--461} (\bibinfo {year} {2006})}\BibitemShut {NoStop}%
\bibitem [{\citenamefont {Haye}, \citenamefont {Isayama},\ and\ \citenamefont
  {Maraschek}(2009)}]{La_Haye_2009}%
  \BibitemOpen
  \bibfield  {author} {\bibinfo {author} {\bibfnamefont {R.~L.}\ \bibnamefont
  {Haye}}, \bibinfo {author} {\bibfnamefont {A.}~\bibnamefont {Isayama}}, \
  and\ \bibinfo {author} {\bibfnamefont {M.}~\bibnamefont {Maraschek}},\
  }\bibfield  {title} {\enquote {\bibinfo {title} {Prospects for stabilization
  of neoclassical tearing modes by electron cyclotron current drive in
  {ITER}},}\ }\href {\doibase 10.1088/0029-5515/49/4/045005} {\bibfield
  {journal} {\bibinfo  {journal} {Nuclear Fusion}\ }\textbf {\bibinfo {volume}
  {49}},\ \bibinfo {pages} {045005} (\bibinfo {year} {2009})}\BibitemShut
  {NoStop}%
\bibitem [{\citenamefont {Reiman}(1983)}]{Reiman_1983}%
  \BibitemOpen
  \bibfield  {author} {\bibinfo {author} {\bibfnamefont {A.~H.}\ \bibnamefont
  {Reiman}},\ }\bibfield  {title} {\enquote {\bibinfo {title} {Suppression of
  magnetic islands by rf driven currents},}\ }\href {\doibase 10.1063/1.864258}
  {\bibfield  {journal} {\bibinfo  {journal} {The Physics of Fluids}\ }\textbf
  {\bibinfo {volume} {26}},\ \bibinfo {pages} {1338--1340} (\bibinfo {year}
  {1983})}\BibitemShut {NoStop}%
\bibitem [{\citenamefont {Hegna}\ and\ \citenamefont
  {Callen}(1997)}]{Hegna_1997}%
  \BibitemOpen
  \bibfield  {author} {\bibinfo {author} {\bibfnamefont {C.~C.}\ \bibnamefont
  {Hegna}}\ and\ \bibinfo {author} {\bibfnamefont {J.~D.}\ \bibnamefont
  {Callen}},\ }\bibfield  {title} {\enquote {\bibinfo {title} {On the
  stabilization of neoclassical magnetohydrodynamic tearing modes using
  localized current drive or heating},}\ }\href {\doibase 10.1063/1.872426}
  {\bibfield  {journal} {\bibinfo  {journal} {Physics of Plasmas}\ }\textbf
  {\bibinfo {volume} {4}},\ \bibinfo {pages} {2940--2946} (\bibinfo {year}
  {1997})}\BibitemShut {NoStop}%
\bibitem [{\citenamefont {Zohm}(1997)}]{Zohm_1997}%
  \BibitemOpen
  \bibfield  {author} {\bibinfo {author} {\bibfnamefont {H.}~\bibnamefont
  {Zohm}},\ }\bibfield  {title} {\enquote {\bibinfo {title} {Stabilization of
  neoclassical tearing modes by electron cyclotron current drive},}\ }\href
  {\doibase 10.1063/1.872487} {\bibfield  {journal} {\bibinfo  {journal}
  {Physics of Plasmas}\ }\textbf {\bibinfo {volume} {4}},\ \bibinfo {pages}
  {3433--3435} (\bibinfo {year} {1997})}\BibitemShut {NoStop}%
\bibitem [{\citenamefont {Yu}\ and\ \citenamefont {Günter}(1998)}]{Yu_1998}%
  \BibitemOpen
  \bibfield  {author} {\bibinfo {author} {\bibfnamefont {Q.}~\bibnamefont
  {Yu}}\ and\ \bibinfo {author} {\bibfnamefont {S.}~\bibnamefont {Günter}},\
  }\bibfield  {title} {\enquote {\bibinfo {title} {On the stabilization of
  neoclassical tearing modes by phased electron cyclotron waves},}\ }\href
  {\doibase 10.1088/0741-3335/40/11/011} {\bibfield  {journal} {\bibinfo
  {journal} {Plasma Physics and Controlled Fusion}\ }\textbf {\bibinfo {volume}
  {40}},\ \bibinfo {pages} {1977--1987} (\bibinfo {year} {1998})}\BibitemShut
  {NoStop}%
\bibitem [{\citenamefont {Harvey}\ and\ \citenamefont
  {Perkins}(2001)}]{Harvey_2001}%
  \BibitemOpen
  \bibfield  {author} {\bibinfo {author} {\bibfnamefont {R.}~\bibnamefont
  {Harvey}}\ and\ \bibinfo {author} {\bibfnamefont {F.}~\bibnamefont
  {Perkins}},\ }\bibfield  {title} {\enquote {\bibinfo {title} {Comparison of
  optimized {ECCD} for different launch locations in a next step tokamak
  reactor plasma},}\ }\href {\doibase 10.1088/0029-5515/41/12/312} {\bibfield
  {journal} {\bibinfo  {journal} {Nuclear Fusion}\ }\textbf {\bibinfo {volume}
  {41}},\ \bibinfo {pages} {1847--1856} (\bibinfo {year} {2001})}\BibitemShut
  {NoStop}%
\bibitem [{\citenamefont {Bernabei}\ \emph {et~al.}(1998)\citenamefont
  {Bernabei}, \citenamefont {Cardinali}, \citenamefont {Giruzzi},\ and\
  \citenamefont {Zabi{\'{e}}go}}]{Bernabei_1998}%
  \BibitemOpen
  \bibfield  {author} {\bibinfo {author} {\bibfnamefont {S.}~\bibnamefont
  {Bernabei}}, \bibinfo {author} {\bibfnamefont {A.}~\bibnamefont {Cardinali}},
  \bibinfo {author} {\bibfnamefont {G.}~\bibnamefont {Giruzzi}}, \ and\
  \bibinfo {author} {\bibfnamefont {M.}~\bibnamefont {Zabi{\'{e}}go}},\
  }\bibfield  {title} {\enquote {\bibinfo {title} {Tearing mode stabilization
  in tokamaks with lower hybrid waves},}\ }\href {\doibase
  10.1088/0029-5515/38/1/307} {\bibfield  {journal} {\bibinfo  {journal}
  {Nuclear Fusion}\ }\textbf {\bibinfo {volume} {38}},\ \bibinfo {pages}
  {87--92} (\bibinfo {year} {1998})}\BibitemShut {NoStop}%
\bibitem [{\citenamefont {Kamendje}\ \emph {et~al.}(2005)\citenamefont
  {Kamendje}, \citenamefont {Kasilov}, \citenamefont {Kernbichler},
  \citenamefont {Pavlenko}, \citenamefont {Poli},\ and\ \citenamefont
  {Heyn}}]{Kamendje_2005}%
  \BibitemOpen
  \bibfield  {author} {\bibinfo {author} {\bibfnamefont {R.}~\bibnamefont
  {Kamendje}}, \bibinfo {author} {\bibfnamefont {S.~V.}\ \bibnamefont
  {Kasilov}}, \bibinfo {author} {\bibfnamefont {W.}~\bibnamefont
  {Kernbichler}}, \bibinfo {author} {\bibfnamefont {I.~V.}\ \bibnamefont
  {Pavlenko}}, \bibinfo {author} {\bibfnamefont {E.}~\bibnamefont {Poli}}, \
  and\ \bibinfo {author} {\bibfnamefont {M.~F.}\ \bibnamefont {Heyn}},\
  }\bibfield  {title} {\enquote {\bibinfo {title} {Modeling of nonlinear
  electron cyclotron resonance heating and current drive in a tokamak},}\
  }\href {\doibase 10.1063/1.1823415} {\bibfield  {journal} {\bibinfo
  {journal} {Physics of Plasmas}\ }\textbf {\bibinfo {volume} {12}},\ \bibinfo
  {pages} {012502} (\bibinfo {year} {2005})}\BibitemShut {NoStop}%
\bibitem [{\citenamefont {La~Haye}(2006)}]{La_Haye_2006b}%
  \BibitemOpen
  \bibfield  {author} {\bibinfo {author} {\bibfnamefont {R.~J.}\ \bibnamefont
  {La~Haye}},\ }\bibfield  {title} {\enquote {\bibinfo {title} {Neoclassical
  tearing modes and their control},}\ }\href {\doibase 10.1063/1.2180747}
  {\bibfield  {journal} {\bibinfo  {journal} {Physics of Plasmas}\ }\textbf
  {\bibinfo {volume} {13}},\ \bibinfo {pages} {055501} (\bibinfo {year}
  {2006})}\BibitemShut {NoStop}%
\bibitem [{\citenamefont {Haye}\ \emph {et~al.}(2008)\citenamefont {Haye},
  \citenamefont {Ferron}, \citenamefont {Humphreys}, \citenamefont {Luce},
  \citenamefont {Petty}, \citenamefont {Prater}, \citenamefont {Strait},\ and\
  \citenamefont {Welander}}]{La_Haye_2008}%
  \BibitemOpen
  \bibfield  {author} {\bibinfo {author} {\bibfnamefont {R.~L.}\ \bibnamefont
  {Haye}}, \bibinfo {author} {\bibfnamefont {J.}~\bibnamefont {Ferron}},
  \bibinfo {author} {\bibfnamefont {D.}~\bibnamefont {Humphreys}}, \bibinfo
  {author} {\bibfnamefont {T.}~\bibnamefont {Luce}}, \bibinfo {author}
  {\bibfnamefont {C.}~\bibnamefont {Petty}}, \bibinfo {author} {\bibfnamefont
  {R.}~\bibnamefont {Prater}}, \bibinfo {author} {\bibfnamefont
  {E.}~\bibnamefont {Strait}}, \ and\ \bibinfo {author} {\bibfnamefont
  {A.}~\bibnamefont {Welander}},\ }\bibfield  {title} {\enquote {\bibinfo
  {title} {Requirements for alignment of electron cyclotron current drive for
  neoclassical tearing mode stabilization in {ITER}},}\ }\href {\doibase
  10.1088/0029-5515/48/5/054004} {\bibfield  {journal} {\bibinfo  {journal}
  {Nuclear Fusion}\ }\textbf {\bibinfo {volume} {48}},\ \bibinfo {pages}
  {054004} (\bibinfo {year} {2008})}\BibitemShut {NoStop}%
\bibitem [{\citenamefont {Sauter}\ \emph {et~al.}(2010)\citenamefont {Sauter},
  \citenamefont {Henderson}, \citenamefont {Ramponi}, \citenamefont {Zohm},\
  and\ \citenamefont {Zucca}}]{Sauter_2010}%
  \BibitemOpen
  \bibfield  {author} {\bibinfo {author} {\bibfnamefont {O.}~\bibnamefont
  {Sauter}}, \bibinfo {author} {\bibfnamefont {M.~A.}\ \bibnamefont
  {Henderson}}, \bibinfo {author} {\bibfnamefont {G.}~\bibnamefont {Ramponi}},
  \bibinfo {author} {\bibfnamefont {H.}~\bibnamefont {Zohm}}, \ and\ \bibinfo
  {author} {\bibfnamefont {C.}~\bibnamefont {Zucca}},\ }\bibfield  {title}
  {\enquote {\bibinfo {title} {On the requirements to control neoclassical
  tearing modes in burning plasmas},}\ }\href {\doibase
  10.1088/0741-3335/52/2/025002} {\bibfield  {journal} {\bibinfo  {journal}
  {Plasma Physics and Controlled Fusion}\ }\textbf {\bibinfo {volume} {52}},\
  \bibinfo {pages} {025002} (\bibinfo {year} {2010})}\BibitemShut {NoStop}%
\bibitem [{\citenamefont {Smolyakov}\ \emph {et~al.}(2013)\citenamefont
  {Smolyakov}, \citenamefont {Poye}, \citenamefont {Agullo}, \citenamefont
  {Benkadda},\ and\ \citenamefont {Garbet}}]{Smolyakov_2013}%
  \BibitemOpen
  \bibfield  {author} {\bibinfo {author} {\bibfnamefont {A.~I.}\ \bibnamefont
  {Smolyakov}}, \bibinfo {author} {\bibfnamefont {A.}~\bibnamefont {Poye}},
  \bibinfo {author} {\bibfnamefont {O.}~\bibnamefont {Agullo}}, \bibinfo
  {author} {\bibfnamefont {S.}~\bibnamefont {Benkadda}}, \ and\ \bibinfo
  {author} {\bibfnamefont {X.}~\bibnamefont {Garbet}},\ }\bibfield  {title}
  {\enquote {\bibinfo {title} {Higher order and asymmetry effects on saturation
  of magnetic islands},}\ }\href {\doibase 10.1063/1.4811383} {\bibfield
  {journal} {\bibinfo  {journal} {Physics of Plasmas}\ }\textbf {\bibinfo
  {volume} {20}},\ \bibinfo {pages} {062506} (\bibinfo {year}
  {2013})}\BibitemShut {NoStop}%
\bibitem [{\citenamefont {Ayten}\ and\ \citenamefont {and}(2014)}]{Ayten_2014}%
  \BibitemOpen
  \bibfield  {author} {\bibinfo {author} {\bibfnamefont {B.}~\bibnamefont
  {Ayten}}\ and\ \bibinfo {author} {\bibfnamefont {E.~W.}\ \bibnamefont
  {and}},\ }\bibfield  {title} {\enquote {\bibinfo {title} {Non-linear effects
  in electron cyclotron current drive applied for the stabilization of
  neoclassical tearing modes},}\ }\href {\doibase
  10.1088/0029-5515/54/7/073001} {\bibfield  {journal} {\bibinfo  {journal}
  {Nuclear Fusion}\ }\textbf {\bibinfo {volume} {54}},\ \bibinfo {pages}
  {073001} (\bibinfo {year} {2014})}\BibitemShut {NoStop}%
\bibitem [{\citenamefont {Borgogno}\ \emph {et~al.}(2014)\citenamefont
  {Borgogno}, \citenamefont {Comisso}, \citenamefont {Grasso},\ and\
  \citenamefont {Lazzaro}}]{Borgogno_2014}%
  \BibitemOpen
  \bibfield  {author} {\bibinfo {author} {\bibfnamefont {D.}~\bibnamefont
  {Borgogno}}, \bibinfo {author} {\bibfnamefont {L.}~\bibnamefont {Comisso}},
  \bibinfo {author} {\bibfnamefont {D.}~\bibnamefont {Grasso}}, \ and\ \bibinfo
  {author} {\bibfnamefont {E.}~\bibnamefont {Lazzaro}},\ }\bibfield  {title}
  {\enquote {\bibinfo {title} {Nonlinear response of magnetic islands to
  localized electron cyclotron current injection},}\ }\href {\doibase
  10.1063/1.4885635} {\bibfield  {journal} {\bibinfo  {journal} {Physics of
  Plasmas}\ }\textbf {\bibinfo {volume} {21}},\ \bibinfo {pages} {060704}
  (\bibinfo {year} {2014})}\BibitemShut {NoStop}%
\bibitem [{\citenamefont {F{\'{e}}vrier}\ \emph {et~al.}(2016)\citenamefont
  {F{\'{e}}vrier}, \citenamefont {Maget}, \citenamefont {Lütjens},
  \citenamefont {Luciani}, \citenamefont {Decker}, \citenamefont {Giruzzi},
  \citenamefont {Reich}, \citenamefont {Beyer}, \citenamefont {Lazzaro},\ and\
  \citenamefont {and}}]{F_vrier_2016}%
  \BibitemOpen
  \bibfield  {author} {\bibinfo {author} {\bibfnamefont {O.}~\bibnamefont
  {F{\'{e}}vrier}}, \bibinfo {author} {\bibfnamefont {P.}~\bibnamefont
  {Maget}}, \bibinfo {author} {\bibfnamefont {H.}~\bibnamefont {Lütjens}},
  \bibinfo {author} {\bibfnamefont {J.~F.}\ \bibnamefont {Luciani}}, \bibinfo
  {author} {\bibfnamefont {J.}~\bibnamefont {Decker}}, \bibinfo {author}
  {\bibfnamefont {G.}~\bibnamefont {Giruzzi}}, \bibinfo {author} {\bibfnamefont
  {M.}~\bibnamefont {Reich}}, \bibinfo {author} {\bibfnamefont
  {P.}~\bibnamefont {Beyer}}, \bibinfo {author} {\bibfnamefont
  {E.}~\bibnamefont {Lazzaro}}, \ and\ \bibinfo {author} {\bibfnamefont
  {S.~N.}\ \bibnamefont {and}},\ }\bibfield  {title} {\enquote {\bibinfo
  {title} {First principles fluid modelling of magnetic island stabilization by
  electron cyclotron current drive ({ECCD})},}\ }\href {\doibase
  10.1088/0741-3335/58/4/045015} {\bibfield  {journal} {\bibinfo  {journal}
  {Plasma Physics and Controlled Fusion}\ }\textbf {\bibinfo {volume} {58}},\
  \bibinfo {pages} {045015} (\bibinfo {year} {2016})}\BibitemShut {NoStop}%
\bibitem [{\citenamefont {Li}\ \emph {et~al.}(2017)\citenamefont {Li},
  \citenamefont {Xiao}, \citenamefont {Lin},\ and\ \citenamefont
  {Wang}}]{Li_2017}%
  \BibitemOpen
  \bibfield  {author} {\bibinfo {author} {\bibfnamefont {J.~C.}\ \bibnamefont
  {Li}}, \bibinfo {author} {\bibfnamefont {C.~J.}\ \bibnamefont {Xiao}},
  \bibinfo {author} {\bibfnamefont {Z.~H.}\ \bibnamefont {Lin}}, \ and\
  \bibinfo {author} {\bibfnamefont {K.~J.}\ \bibnamefont {Wang}},\ }\bibfield
  {title} {\enquote {\bibinfo {title} {Effects of electron cyclotron current
  drive on magnetic islands in tokamak plasmas},}\ }\href {\doibase
  10.1063/1.4996021} {\bibfield  {journal} {\bibinfo  {journal} {Physics of
  Plasmas}\ }\textbf {\bibinfo {volume} {24}},\ \bibinfo {pages} {082508}
  (\bibinfo {year} {2017})}\BibitemShut {NoStop}%
\bibitem [{\citenamefont {Grasso}\ \emph {et~al.}(2016)\citenamefont {Grasso},
  \citenamefont {Lazzaro}, \citenamefont {Borgogno},\ and\ \citenamefont
  {Comisso}}]{Grasso_2016}%
  \BibitemOpen
  \bibfield  {author} {\bibinfo {author} {\bibfnamefont {D.}~\bibnamefont
  {Grasso}}, \bibinfo {author} {\bibfnamefont {E.}~\bibnamefont {Lazzaro}},
  \bibinfo {author} {\bibfnamefont {D.}~\bibnamefont {Borgogno}}, \ and\
  \bibinfo {author} {\bibfnamefont {L.}~\bibnamefont {Comisso}},\ }\bibfield
  {title} {\enquote {\bibinfo {title} {Open problems of magnetic island control
  by electron cyclotron current drive},}\ }\href {\doibase
  10.1017/S0022377816001045} {\bibfield  {journal} {\bibinfo  {journal}
  {Journal of Plasma Physics}\ }\textbf {\bibinfo {volume} {82}},\ \bibinfo
  {pages} {595820603} (\bibinfo {year} {2016})}\BibitemShut {NoStop}%
\bibitem [{\citenamefont {Zohm}\ \emph {et~al.}(1999)\citenamefont {Zohm},
  \citenamefont {Gantenbein}, \citenamefont {Giruzzi}, \citenamefont {Günter},
  \citenamefont {Leuterer}, \citenamefont {Maraschek}, \citenamefont {Meskat},
  \citenamefont {Peeters}, \citenamefont {Suttrop}, \citenamefont {Wagner},
  \citenamefont {Zabi{\'{e}}go}, \citenamefont {Team},\ and\ \citenamefont
  {Group}}]{Zohm_1999}%
  \BibitemOpen
  \bibfield  {author} {\bibinfo {author} {\bibfnamefont {H.}~\bibnamefont
  {Zohm}}, \bibinfo {author} {\bibfnamefont {G.}~\bibnamefont {Gantenbein}},
  \bibinfo {author} {\bibfnamefont {G.}~\bibnamefont {Giruzzi}}, \bibinfo
  {author} {\bibfnamefont {S.}~\bibnamefont {Günter}}, \bibinfo {author}
  {\bibfnamefont {F.}~\bibnamefont {Leuterer}}, \bibinfo {author}
  {\bibfnamefont {M.}~\bibnamefont {Maraschek}}, \bibinfo {author}
  {\bibfnamefont {J.}~\bibnamefont {Meskat}}, \bibinfo {author} {\bibfnamefont
  {A.}~\bibnamefont {Peeters}}, \bibinfo {author} {\bibfnamefont
  {W.}~\bibnamefont {Suttrop}}, \bibinfo {author} {\bibfnamefont
  {D.}~\bibnamefont {Wagner}}, \bibinfo {author} {\bibfnamefont
  {M.}~\bibnamefont {Zabi{\'{e}}go}}, \bibinfo {author} {\bibfnamefont {A.~U.}\
  \bibnamefont {Team}}, \ and\ \bibinfo {author} {\bibfnamefont
  {E.}~\bibnamefont {Group}},\ }\bibfield  {title} {\enquote {\bibinfo {title}
  {Experiments on neoclassical tearing mode stabilization by {ECCD} in {ASDEX}
  upgrade},}\ }\href {\doibase 10.1088/0029-5515/39/5/101} {\bibfield
  {journal} {\bibinfo  {journal} {Nuclear Fusion}\ }\textbf {\bibinfo {volume}
  {39}},\ \bibinfo {pages} {577--580} (\bibinfo {year} {1999})}\BibitemShut
  {NoStop}%
\bibitem [{\citenamefont {Prater}\ \emph {et~al.}(2003)\citenamefont {Prater},
  \citenamefont {Haye}, \citenamefont {Lohr}, \citenamefont {Luce},
  \citenamefont {Petty}, \citenamefont {Ferron}, \citenamefont {Humphreys},
  \citenamefont {Strait}, \citenamefont {Perkins},\ and\ \citenamefont
  {Harvey}}]{Prater_2003}%
  \BibitemOpen
  \bibfield  {author} {\bibinfo {author} {\bibfnamefont {R.}~\bibnamefont
  {Prater}}, \bibinfo {author} {\bibfnamefont {R.~L.}\ \bibnamefont {Haye}},
  \bibinfo {author} {\bibfnamefont {J.}~\bibnamefont {Lohr}}, \bibinfo {author}
  {\bibfnamefont {T.}~\bibnamefont {Luce}}, \bibinfo {author} {\bibfnamefont
  {C.}~\bibnamefont {Petty}}, \bibinfo {author} {\bibfnamefont
  {J.}~\bibnamefont {Ferron}}, \bibinfo {author} {\bibfnamefont
  {D.}~\bibnamefont {Humphreys}}, \bibinfo {author} {\bibfnamefont
  {E.}~\bibnamefont {Strait}}, \bibinfo {author} {\bibfnamefont
  {F.}~\bibnamefont {Perkins}}, \ and\ \bibinfo {author} {\bibfnamefont
  {R.}~\bibnamefont {Harvey}},\ }\bibfield  {title} {\enquote {\bibinfo {title}
  {Discharge improvement through control of neoclassical tearing modes by
  localized {ECCD} in {DIII-D}},}\ }\href {\doibase
  10.1088/0029-5515/43/10/014} {\bibfield  {journal} {\bibinfo  {journal}
  {Nuclear Fusion}\ }\textbf {\bibinfo {volume} {43}},\ \bibinfo {pages}
  {1128--1134} (\bibinfo {year} {2003})}\BibitemShut {NoStop}%
\bibitem [{\citenamefont {Warrick}\ \emph {et~al.}(2000)\citenamefont
  {Warrick}, \citenamefont {Buttery}, \citenamefont {Cunningham}, \citenamefont
  {Fielding}, \citenamefont {Hender}, \citenamefont {Lloyd}, \citenamefont
  {Morris}, \citenamefont {O'Brien}, \citenamefont {Pinfold}, \citenamefont
  {Stammers}, \citenamefont {Valovic}, \citenamefont {Walsh}, \citenamefont
  {Wilson}, \citenamefont {COMPASS-D},\ and\ \citenamefont
  {teams}}]{Warrick_2000}%
  \BibitemOpen
  \bibfield  {author} {\bibinfo {author} {\bibfnamefont {C.~D.}\ \bibnamefont
  {Warrick}}, \bibinfo {author} {\bibfnamefont {R.~J.}\ \bibnamefont
  {Buttery}}, \bibinfo {author} {\bibfnamefont {G.}~\bibnamefont {Cunningham}},
  \bibinfo {author} {\bibfnamefont {S.~J.}\ \bibnamefont {Fielding}}, \bibinfo
  {author} {\bibfnamefont {T.~C.}\ \bibnamefont {Hender}}, \bibinfo {author}
  {\bibfnamefont {B.}~\bibnamefont {Lloyd}}, \bibinfo {author} {\bibfnamefont
  {A.~W.}\ \bibnamefont {Morris}}, \bibinfo {author} {\bibfnamefont {M.~R.}\
  \bibnamefont {O'Brien}}, \bibinfo {author} {\bibfnamefont {T.}~\bibnamefont
  {Pinfold}}, \bibinfo {author} {\bibfnamefont {K.}~\bibnamefont {Stammers}},
  \bibinfo {author} {\bibfnamefont {M.}~\bibnamefont {Valovic}}, \bibinfo
  {author} {\bibfnamefont {M.}~\bibnamefont {Walsh}}, \bibinfo {author}
  {\bibfnamefont {H.~R.}\ \bibnamefont {Wilson}}, \bibinfo {author}
  {\bibnamefont {COMPASS-D}}, \ and\ \bibinfo {author} {\bibfnamefont
  {R.}~\bibnamefont {teams}},\ }\bibfield  {title} {\enquote {\bibinfo {title}
  {Complete stabilization of neoclassical tearing modes with lower hybrid
  current drive on {COMPASS-D}},}\ }\href {\doibase 10.1103/PhysRevLett.85.574}
  {\bibfield  {journal} {\bibinfo  {journal} {Phys. Rev. Lett.}\ }\textbf
  {\bibinfo {volume} {85}},\ \bibinfo {pages} {574--577} (\bibinfo {year}
  {2000})}\BibitemShut {NoStop}%
\bibitem [{\citenamefont {Gantenbein}\ \emph {et~al.}(2000)\citenamefont
  {Gantenbein}, \citenamefont {Zohm}, \citenamefont {Giruzzi}, \citenamefont
  {G\"unter}, \citenamefont {Leuterer}, \citenamefont {Maraschek},
  \citenamefont {Meskat}, \citenamefont {Yu}, \citenamefont {Team},\ and\
  \citenamefont {(AUG)}}]{Gantenbein_2000}%
  \BibitemOpen
  \bibfield  {author} {\bibinfo {author} {\bibfnamefont {G.}~\bibnamefont
  {Gantenbein}}, \bibinfo {author} {\bibfnamefont {H.}~\bibnamefont {Zohm}},
  \bibinfo {author} {\bibfnamefont {G.}~\bibnamefont {Giruzzi}}, \bibinfo
  {author} {\bibfnamefont {S.}~\bibnamefont {G\"unter}}, \bibinfo {author}
  {\bibfnamefont {F.}~\bibnamefont {Leuterer}}, \bibinfo {author}
  {\bibfnamefont {M.}~\bibnamefont {Maraschek}}, \bibinfo {author}
  {\bibfnamefont {J.}~\bibnamefont {Meskat}}, \bibinfo {author} {\bibfnamefont
  {Q.}~\bibnamefont {Yu}}, \bibinfo {author} {\bibfnamefont {A.~U.}\
  \bibnamefont {Team}}, \ and\ \bibinfo {author} {\bibfnamefont {E.-G.}\
  \bibnamefont {(AUG)}},\ }\bibfield  {title} {\enquote {\bibinfo {title}
  {Complete suppression of neoclassical tearing modes with current drive at the
  electron-cyclotron-resonance frequency in {ASDEX} upgrade tokamak},}\ }\href
  {\doibase 10.1103/PhysRevLett.85.1242} {\bibfield  {journal} {\bibinfo
  {journal} {Phys. Rev. Lett.}\ }\textbf {\bibinfo {volume} {85}},\ \bibinfo
  {pages} {1242--1245} (\bibinfo {year} {2000})}\BibitemShut {NoStop}%
\bibitem [{\citenamefont {Zohm}\ \emph {et~al.}(2001)\citenamefont {Zohm},
  \citenamefont {Gantenbein}, \citenamefont {Gude}, \citenamefont {Günter},
  \citenamefont {Leuterer}, \citenamefont {Maraschek}, \citenamefont {Meskat},
  \citenamefont {Suttrop}, \citenamefont {Yu}, \citenamefont {Team},\ and\
  \citenamefont {(AUG)}}]{Zohm_2001}%
  \BibitemOpen
  \bibfield  {author} {\bibinfo {author} {\bibfnamefont {H.}~\bibnamefont
  {Zohm}}, \bibinfo {author} {\bibfnamefont {G.}~\bibnamefont {Gantenbein}},
  \bibinfo {author} {\bibfnamefont {A.}~\bibnamefont {Gude}}, \bibinfo {author}
  {\bibfnamefont {S.}~\bibnamefont {Günter}}, \bibinfo {author} {\bibfnamefont
  {F.}~\bibnamefont {Leuterer}}, \bibinfo {author} {\bibfnamefont
  {M.}~\bibnamefont {Maraschek}}, \bibinfo {author} {\bibfnamefont
  {J.}~\bibnamefont {Meskat}}, \bibinfo {author} {\bibfnamefont
  {W.}~\bibnamefont {Suttrop}}, \bibinfo {author} {\bibfnamefont
  {Q.}~\bibnamefont {Yu}}, \bibinfo {author} {\bibfnamefont {A.~U.}\
  \bibnamefont {Team}}, \ and\ \bibinfo {author} {\bibfnamefont {E.~G.}\
  \bibnamefont {(AUG)}},\ }\bibfield  {title} {\enquote {\bibinfo {title} {The
  physics of neoclassical tearing modes and their stabilization by {ECCD} in
  {ASDEX} upgrade},}\ }\href {\doibase 10.1088/0029-5515/41/2/306} {\bibfield
  {journal} {\bibinfo  {journal} {Nuclear Fusion}\ }\textbf {\bibinfo {volume}
  {41}},\ \bibinfo {pages} {197--202} (\bibinfo {year} {2001})}\BibitemShut
  {NoStop}%
\bibitem [{\citenamefont {Isayama}\ \emph {et~al.}(2000)\citenamefont
  {Isayama}, \citenamefont {Kamada}, \citenamefont {Ide}, \citenamefont
  {Hamamatsu}, \citenamefont {Oikawa}, \citenamefont {Suzuki}, \citenamefont
  {Neyatani}, \citenamefont {Ozeki}, \citenamefont {Ikeda},\ and\ \citenamefont
  {and}}]{Isayama_2000}%
  \BibitemOpen
  \bibfield  {author} {\bibinfo {author} {\bibfnamefont {A.}~\bibnamefont
  {Isayama}}, \bibinfo {author} {\bibfnamefont {Y.}~\bibnamefont {Kamada}},
  \bibinfo {author} {\bibfnamefont {S.}~\bibnamefont {Ide}}, \bibinfo {author}
  {\bibfnamefont {K.}~\bibnamefont {Hamamatsu}}, \bibinfo {author}
  {\bibfnamefont {T.}~\bibnamefont {Oikawa}}, \bibinfo {author} {\bibfnamefont
  {T.}~\bibnamefont {Suzuki}}, \bibinfo {author} {\bibfnamefont
  {Y.}~\bibnamefont {Neyatani}}, \bibinfo {author} {\bibfnamefont
  {T.}~\bibnamefont {Ozeki}}, \bibinfo {author} {\bibfnamefont
  {Y.}~\bibnamefont {Ikeda}}, \ and\ \bibinfo {author} {\bibfnamefont {K.~K.}\
  \bibnamefont {and}},\ }\bibfield  {title} {\enquote {\bibinfo {title}
  {Complete stabilization of a tearing mode in steady state high- ph-mode
  discharges by the first harmonic electron cyclotron heating/current drive on
  {JT-60U}},}\ }\href {\doibase 10.1088/0741-3335/42/12/102} {\bibfield
  {journal} {\bibinfo  {journal} {Plasma Physics and Controlled Fusion}\
  }\textbf {\bibinfo {volume} {42}},\ \bibinfo {pages} {L37--L45} (\bibinfo
  {year} {2000})}\BibitemShut {NoStop}%
\bibitem [{\citenamefont {La~Haye}\ \emph {et~al.}(2002)\citenamefont
  {La~Haye}, \citenamefont {Günter}, \citenamefont {Humphreys}, \citenamefont
  {Lohr}, \citenamefont {Luce}, \citenamefont {Maraschek}, \citenamefont
  {Petty}, \citenamefont {Prater}, \citenamefont {Scoville},\ and\
  \citenamefont {Strait}}]{La_Haye_2002}%
  \BibitemOpen
  \bibfield  {author} {\bibinfo {author} {\bibfnamefont {R.~J.}\ \bibnamefont
  {La~Haye}}, \bibinfo {author} {\bibfnamefont {S.}~\bibnamefont {Günter}},
  \bibinfo {author} {\bibfnamefont {D.~A.}\ \bibnamefont {Humphreys}}, \bibinfo
  {author} {\bibfnamefont {J.}~\bibnamefont {Lohr}}, \bibinfo {author}
  {\bibfnamefont {T.~C.}\ \bibnamefont {Luce}}, \bibinfo {author}
  {\bibfnamefont {M.~E.}\ \bibnamefont {Maraschek}}, \bibinfo {author}
  {\bibfnamefont {C.~C.}\ \bibnamefont {Petty}}, \bibinfo {author}
  {\bibfnamefont {R.}~\bibnamefont {Prater}}, \bibinfo {author} {\bibfnamefont
  {J.~T.}\ \bibnamefont {Scoville}}, \ and\ \bibinfo {author} {\bibfnamefont
  {E.~J.}\ \bibnamefont {Strait}},\ }\bibfield  {title} {\enquote {\bibinfo
  {title} {Control of neoclassical tearing modes in {DIII–D}},}\ }\href
  {\doibase 10.1063/1.1456066} {\bibfield  {journal} {\bibinfo  {journal}
  {Physics of Plasmas}\ }\textbf {\bibinfo {volume} {9}},\ \bibinfo {pages}
  {2051--2060} (\bibinfo {year} {2002})}\BibitemShut {NoStop}%
\bibitem [{\citenamefont {Petty}\ \emph {et~al.}(2004)\citenamefont {Petty},
  \citenamefont {Haye}, \citenamefont {Luce}, \citenamefont {Humphreys},
  \citenamefont {Hyatt}, \citenamefont {Lohr}, \citenamefont {Prater},
  \citenamefont {Strait},\ and\ \citenamefont {Wade}}]{Petty_2004}%
  \BibitemOpen
  \bibfield  {author} {\bibinfo {author} {\bibfnamefont {C.}~\bibnamefont
  {Petty}}, \bibinfo {author} {\bibfnamefont {R.~L.}\ \bibnamefont {Haye}},
  \bibinfo {author} {\bibfnamefont {T.}~\bibnamefont {Luce}}, \bibinfo {author}
  {\bibfnamefont {D.}~\bibnamefont {Humphreys}}, \bibinfo {author}
  {\bibfnamefont {A.}~\bibnamefont {Hyatt}}, \bibinfo {author} {\bibfnamefont
  {J.}~\bibnamefont {Lohr}}, \bibinfo {author} {\bibfnamefont {R.}~\bibnamefont
  {Prater}}, \bibinfo {author} {\bibfnamefont {E.}~\bibnamefont {Strait}}, \
  and\ \bibinfo {author} {\bibfnamefont {M.}~\bibnamefont {Wade}},\ }\bibfield
  {title} {\enquote {\bibinfo {title} {Complete suppression of the m= 2/n= 1
  neoclassical tearing mode using electron cyclotron current drive in
  {DIII-D}},}\ }\href {\doibase 10.1088/0029-5515/44/2/004} {\bibfield
  {journal} {\bibinfo  {journal} {Nuclear Fusion}\ }\textbf {\bibinfo {volume}
  {44}},\ \bibinfo {pages} {243--251} (\bibinfo {year} {2004})}\BibitemShut
  {NoStop}%
\bibitem [{\citenamefont {Volpe}\ \emph {et~al.}(2015)\citenamefont {Volpe},
  \citenamefont {Hyatt}, \citenamefont {La~Haye}, \citenamefont {Lanctot},
  \citenamefont {Lohr}, \citenamefont {Prater}, \citenamefont {Strait},\ and\
  \citenamefont {Welander}}]{Volpe_2015}%
  \BibitemOpen
  \bibfield  {author} {\bibinfo {author} {\bibfnamefont {F.~A.}\ \bibnamefont
  {Volpe}}, \bibinfo {author} {\bibfnamefont {A.}~\bibnamefont {Hyatt}},
  \bibinfo {author} {\bibfnamefont {R.~J.}\ \bibnamefont {La~Haye}}, \bibinfo
  {author} {\bibfnamefont {M.~J.}\ \bibnamefont {Lanctot}}, \bibinfo {author}
  {\bibfnamefont {J.}~\bibnamefont {Lohr}}, \bibinfo {author} {\bibfnamefont
  {R.}~\bibnamefont {Prater}}, \bibinfo {author} {\bibfnamefont {E.~J.}\
  \bibnamefont {Strait}}, \ and\ \bibinfo {author} {\bibfnamefont
  {A.}~\bibnamefont {Welander}},\ }\bibfield  {title} {\enquote {\bibinfo
  {title} {Avoiding tokamak disruptions by applying static magnetic fields that
  align locked modes with stabilizing wave-driven currents},}\ }\href {\doibase
  10.1103/PhysRevLett.115.175002} {\bibfield  {journal} {\bibinfo  {journal}
  {Phys. Rev. Lett.}\ }\textbf {\bibinfo {volume} {115}},\ \bibinfo {pages}
  {175002} (\bibinfo {year} {2015})}\BibitemShut {NoStop}%
\bibitem [{\citenamefont {Reiman}\ and\ \citenamefont
  {Fisch}(2018)}]{Reiman_2018}%
  \BibitemOpen
  \bibfield  {author} {\bibinfo {author} {\bibfnamefont {A.~H.}\ \bibnamefont
  {Reiman}}\ and\ \bibinfo {author} {\bibfnamefont {N.~J.}\ \bibnamefont
  {Fisch}},\ }\bibfield  {title} {\enquote {\bibinfo {title} {Suppression of
  tearing modes by radio frequency current condensation},}\ }\href {\doibase
  10.1103/PhysRevLett.121.225001} {\bibfield  {journal} {\bibinfo  {journal}
  {Phys. Rev. Lett.}\ }\textbf {\bibinfo {volume} {121}},\ \bibinfo {pages}
  {225001} (\bibinfo {year} {2018})}\BibitemShut {NoStop}%
\bibitem [{\citenamefont {Rodríguez}, \citenamefont {Reiman},\ and\
  \citenamefont {Fisch}(2019)}]{Rodriguez_2019}%
  \BibitemOpen
  \bibfield  {author} {\bibinfo {author} {\bibfnamefont {E.}~\bibnamefont
  {Rodríguez}}, \bibinfo {author} {\bibfnamefont {A.~H.}\ \bibnamefont
  {Reiman}}, \ and\ \bibinfo {author} {\bibfnamefont {N.~J.}\ \bibnamefont
  {Fisch}},\ }\bibfield  {title} {\enquote {\bibinfo {title} {Rf current
  condensation in magnetic islands and associated hysteresis phenomena},}\
  }\href {\doibase 10.1063/1.5118424} {\bibfield  {journal} {\bibinfo
  {journal} {Physics of Plasmas}\ }\textbf {\bibinfo {volume} {26}},\ \bibinfo
  {pages} {092511} (\bibinfo {year} {2019})}\BibitemShut {NoStop}%
\bibitem [{\citenamefont {Rodríguez}, \citenamefont {Reiman},\ and\
  \citenamefont {Fisch}(2020)}]{eduardo}%
  \BibitemOpen
  \bibfield  {author} {\bibinfo {author} {\bibfnamefont {E.}~\bibnamefont
  {Rodríguez}}, \bibinfo {author} {\bibfnamefont {A.~H.}\ \bibnamefont
  {Reiman}}, \ and\ \bibinfo {author} {\bibfnamefont {N.~J.}\ \bibnamefont
  {Fisch}},\ }\bibfield  {title} {\enquote {\bibinfo {title} {Rf current
  condensation in the presence of turbulent enhanced transport},}\ }\href
  {\doibase 10.1063/5.0001881} {\bibfield  {journal} {\bibinfo  {journal}
  {Physics of Plasmas}\ }\textbf {\bibinfo {volume} {27}},\ \bibinfo {pages}
  {042306} (\bibinfo {year} {2020})}\BibitemShut {NoStop}%
\bibitem [{\citenamefont {Jin}, \citenamefont {Fisch},\ and\ \citenamefont
  {Reiman}(2020)}]{Jin_2020}%
  \BibitemOpen
  \bibfield  {author} {\bibinfo {author} {\bibfnamefont {S.}~\bibnamefont
  {Jin}}, \bibinfo {author} {\bibfnamefont {N.~J.}\ \bibnamefont {Fisch}}, \
  and\ \bibinfo {author} {\bibfnamefont {A.~H.}\ \bibnamefont {Reiman}},\
  }\bibfield  {title} {\enquote {\bibinfo {title} {Pulsed rf schemes for
  tearing mode stabilization},}\ }\href {\doibase 10.1063/5.0007861} {\bibfield
   {journal} {\bibinfo  {journal} {Physics of Plasmas}\ }\textbf {\bibinfo
  {volume} {27}},\ \bibinfo {pages} {062508} (\bibinfo {year}
  {2020})}\BibitemShut {NoStop}%
\bibitem [{\citenamefont {Nies}\ \emph {et~al.}(2020)\citenamefont {Nies},
  \citenamefont {Reiman}, \citenamefont {Rodriguez}, \citenamefont {Bertelli},\
  and\ \citenamefont {Fisch}}]{nies_2020}%
  \BibitemOpen
  \bibfield  {author} {\bibinfo {author} {\bibfnamefont {R.}~\bibnamefont
  {Nies}}, \bibinfo {author} {\bibfnamefont {A.~H.}\ \bibnamefont {Reiman}},
  \bibinfo {author} {\bibfnamefont {E.}~\bibnamefont {Rodriguez}}, \bibinfo
  {author} {\bibfnamefont {N.}~\bibnamefont {Bertelli}}, \ and\ \bibinfo
  {author} {\bibfnamefont {N.~J.}\ \bibnamefont {Fisch}},\ }\bibfield  {title}
  {\enquote {\bibinfo {title} {Calculating rf current condensation with
  consistent ray-tracing and island heating},}\ }\href {\doibase
  10.1063/5.0013573} {\bibfield  {journal} {\bibinfo  {journal} {Physics of
  Plasmas}\ }\textbf {\bibinfo {volume} {27}},\ \bibinfo {pages} {092503}
  (\bibinfo {year} {2020})}\BibitemShut {NoStop}%
\bibitem [{\citenamefont {Spakman}\ \emph {et~al.}(2008)\citenamefont
  {Spakman}, \citenamefont {Hogeweij}, \citenamefont {Jaspers}, \citenamefont
  {Schüller}, \citenamefont {Westerhof}, \citenamefont {Boom}, \citenamefont
  {Classen}, \citenamefont {Delabie}, \citenamefont {Domier}, \citenamefont
  {Donn{\'{e}}}, \citenamefont {Kantor}, \citenamefont {Krämer-Flecken},
  \citenamefont {Liang}, \citenamefont {Luhmann}, \citenamefont {Park},
  \citenamefont {van~de Pol}, \citenamefont {Schmitz},\ and\ \citenamefont
  {and}}]{Spakman_2008}%
  \BibitemOpen
  \bibfield  {author} {\bibinfo {author} {\bibfnamefont {G.}~\bibnamefont
  {Spakman}}, \bibinfo {author} {\bibfnamefont {G.}~\bibnamefont {Hogeweij}},
  \bibinfo {author} {\bibfnamefont {R.}~\bibnamefont {Jaspers}}, \bibinfo
  {author} {\bibfnamefont {F.}~\bibnamefont {Schüller}}, \bibinfo {author}
  {\bibfnamefont {E.}~\bibnamefont {Westerhof}}, \bibinfo {author}
  {\bibfnamefont {J.}~\bibnamefont {Boom}}, \bibinfo {author} {\bibfnamefont
  {I.}~\bibnamefont {Classen}}, \bibinfo {author} {\bibfnamefont
  {E.}~\bibnamefont {Delabie}}, \bibinfo {author} {\bibfnamefont
  {C.}~\bibnamefont {Domier}}, \bibinfo {author} {\bibfnamefont
  {A.}~\bibnamefont {Donn{\'{e}}}}, \bibinfo {author} {\bibfnamefont
  {M.}~\bibnamefont {Kantor}}, \bibinfo {author} {\bibfnamefont
  {A.}~\bibnamefont {Krämer-Flecken}}, \bibinfo {author} {\bibfnamefont
  {Y.}~\bibnamefont {Liang}}, \bibinfo {author} {\bibfnamefont
  {N.}~\bibnamefont {Luhmann}}, \bibinfo {author} {\bibfnamefont
  {H.}~\bibnamefont {Park}}, \bibinfo {author} {\bibfnamefont {M.}~\bibnamefont
  {van~de Pol}}, \bibinfo {author} {\bibfnamefont {O.}~\bibnamefont {Schmitz}},
  \ and\ \bibinfo {author} {\bibfnamefont {J.~O.}\ \bibnamefont {and}},\
  }\bibfield  {title} {\enquote {\bibinfo {title} {Heat pulse propagation
  studies around magnetic islands induced by the dynamic ergodic divertor in
  {TEXTOR}},}\ }\href {\doibase 10.1088/0029-5515/48/11/115005} {\bibfield
  {journal} {\bibinfo  {journal} {Nuclear Fusion}\ }\textbf {\bibinfo {volume}
  {48}},\ \bibinfo {pages} {115005} (\bibinfo {year} {2008})}\BibitemShut
  {NoStop}%
\bibitem [{\citenamefont {Inagaki}\ \emph {et~al.}(2004)\citenamefont
  {Inagaki}, \citenamefont {Tamura}, \citenamefont {Ida}, \citenamefont
  {Nagayama}, \citenamefont {Kawahata}, \citenamefont {Sudo}, \citenamefont
  {Morisaki}, \citenamefont {Tanaka},\ and\ \citenamefont
  {Tokuzawa}}]{Inagaki_2004}%
  \BibitemOpen
  \bibfield  {author} {\bibinfo {author} {\bibfnamefont {S.}~\bibnamefont
  {Inagaki}}, \bibinfo {author} {\bibfnamefont {N.}~\bibnamefont {Tamura}},
  \bibinfo {author} {\bibfnamefont {K.}~\bibnamefont {Ida}}, \bibinfo {author}
  {\bibfnamefont {Y.}~\bibnamefont {Nagayama}}, \bibinfo {author}
  {\bibfnamefont {K.}~\bibnamefont {Kawahata}}, \bibinfo {author}
  {\bibfnamefont {S.}~\bibnamefont {Sudo}}, \bibinfo {author} {\bibfnamefont
  {T.}~\bibnamefont {Morisaki}}, \bibinfo {author} {\bibfnamefont
  {K.}~\bibnamefont {Tanaka}}, \ and\ \bibinfo {author} {\bibfnamefont
  {T.}~\bibnamefont {Tokuzawa}} (\bibinfo {collaboration} {the LHD Experimental
  Group}),\ }\bibfield  {title} {\enquote {\bibinfo {title} {Observation of
  reduced heat transport inside the magnetic island o point in the large
  helical device},}\ }\href {\doibase 10.1103/PhysRevLett.92.055002} {\bibfield
   {journal} {\bibinfo  {journal} {Phys. Rev. Lett.}\ }\textbf {\bibinfo
  {volume} {92}},\ \bibinfo {pages} {055002} (\bibinfo {year}
  {2004})}\BibitemShut {NoStop}%
\bibitem [{\citenamefont {Bardóczi}\ \emph {et~al.}(2016)\citenamefont
  {Bardóczi}, \citenamefont {Rhodes}, \citenamefont {Carter}, \citenamefont
  {Crocker}, \citenamefont {Peebles},\ and\ \citenamefont
  {Grierson}}]{Bardoczi_2016}%
  \BibitemOpen
  \bibfield  {author} {\bibinfo {author} {\bibfnamefont {L.}~\bibnamefont
  {Bardóczi}}, \bibinfo {author} {\bibfnamefont {T.~L.}\ \bibnamefont
  {Rhodes}}, \bibinfo {author} {\bibfnamefont {T.~A.}\ \bibnamefont {Carter}},
  \bibinfo {author} {\bibfnamefont {N.~A.}\ \bibnamefont {Crocker}}, \bibinfo
  {author} {\bibfnamefont {W.~A.}\ \bibnamefont {Peebles}}, \ and\ \bibinfo
  {author} {\bibfnamefont {B.~A.}\ \bibnamefont {Grierson}},\ }\bibfield
  {title} {\enquote {\bibinfo {title} {Non-perturbative measurement of
  cross-field thermal diffusivity reduction at the o-point of 2/1 neoclassical
  tearing mode islands in the {DIII-D} tokamak},}\ }\href {\doibase
  10.1063/1.4948560} {\bibfield  {journal} {\bibinfo  {journal} {Physics of
  Plasmas}\ }\textbf {\bibinfo {volume} {23}},\ \bibinfo {pages} {052507}
  (\bibinfo {year} {2016})}\BibitemShut {NoStop}%
\bibitem [{\citenamefont {Ida}\ \emph {et~al.}(2012)\citenamefont {Ida},
  \citenamefont {Kamiya}, \citenamefont {Isayama},\ and\ \citenamefont
  {Sakamoto}}]{Ida_2012}%
  \BibitemOpen
  \bibfield  {author} {\bibinfo {author} {\bibfnamefont {K.}~\bibnamefont
  {Ida}}, \bibinfo {author} {\bibfnamefont {K.}~\bibnamefont {Kamiya}},
  \bibinfo {author} {\bibfnamefont {A.}~\bibnamefont {Isayama}}, \ and\
  \bibinfo {author} {\bibfnamefont {Y.}~\bibnamefont {Sakamoto}} (\bibinfo
  {collaboration} {JT-60 Team}),\ }\bibfield  {title} {\enquote {\bibinfo
  {title} {Reduction of ion thermal diffusivity inside a magnetic island in
  {JT-60U} tokamak plasma},}\ }\href {\doibase 10.1103/PhysRevLett.109.065001}
  {\bibfield  {journal} {\bibinfo  {journal} {Phys. Rev. Lett.}\ }\textbf
  {\bibinfo {volume} {109}},\ \bibinfo {pages} {065001} (\bibinfo {year}
  {2012})}\BibitemShut {NoStop}%
\bibitem [{\citenamefont {Westerhof}\ \emph {et~al.}(2007)\citenamefont
  {Westerhof}, \citenamefont {Lazaros}, \citenamefont {Farshi}, \citenamefont
  {de~Baar}, \citenamefont {de~Bock}, \citenamefont {Classen}, \citenamefont
  {Jaspers}, \citenamefont {Hogeweij}, \citenamefont {Koslowski}, \citenamefont
  {Krämer-Flecken}, \citenamefont {Liang}, \citenamefont {Cardozo},\ and\
  \citenamefont {Zimmermann}}]{Westerhof_2007}%
  \BibitemOpen
  \bibfield  {author} {\bibinfo {author} {\bibfnamefont {E.}~\bibnamefont
  {Westerhof}}, \bibinfo {author} {\bibfnamefont {A.}~\bibnamefont {Lazaros}},
  \bibinfo {author} {\bibfnamefont {E.}~\bibnamefont {Farshi}}, \bibinfo
  {author} {\bibfnamefont {M.}~\bibnamefont {de~Baar}}, \bibinfo {author}
  {\bibfnamefont {M.}~\bibnamefont {de~Bock}}, \bibinfo {author} {\bibfnamefont
  {I.}~\bibnamefont {Classen}}, \bibinfo {author} {\bibfnamefont
  {R.}~\bibnamefont {Jaspers}}, \bibinfo {author} {\bibfnamefont
  {G.}~\bibnamefont {Hogeweij}}, \bibinfo {author} {\bibfnamefont
  {H.}~\bibnamefont {Koslowski}}, \bibinfo {author} {\bibfnamefont
  {A.}~\bibnamefont {Krämer-Flecken}}, \bibinfo {author} {\bibfnamefont
  {Y.}~\bibnamefont {Liang}}, \bibinfo {author} {\bibfnamefont {N.~L.}\
  \bibnamefont {Cardozo}}, \ and\ \bibinfo {author} {\bibfnamefont
  {O.}~\bibnamefont {Zimmermann}},\ }\bibfield  {title} {\enquote {\bibinfo
  {title} {Tearing mode stabilization by electron cyclotron resonance heating
  demonstrated in the {TEXTOR} tokamak and the implication for {ITER}},}\
  }\href {\doibase 10.1088/0029-5515/47/2/003} {\bibfield  {journal} {\bibinfo
  {journal} {Nuclear Fusion}\ }\textbf {\bibinfo {volume} {47}},\ \bibinfo
  {pages} {85--90} (\bibinfo {year} {2007})}\BibitemShut {NoStop}%
\bibitem [{\citenamefont {Fisch}(1978)}]{fisch_1978}%
  \BibitemOpen
  \bibfield  {author} {\bibinfo {author} {\bibfnamefont {N.~J.}\ \bibnamefont
  {Fisch}},\ }\bibfield  {title} {\enquote {\bibinfo {title} {Confining a
  tokamak plasma with rf-driven currents},}\ }\href {\doibase
  10.1103/PhysRevLett.41.873} {\bibfield  {journal} {\bibinfo  {journal} {Phys.
  Rev. Lett.}\ }\textbf {\bibinfo {volume} {41}},\ \bibinfo {pages} {873--876}
  (\bibinfo {year} {1978})}\BibitemShut {NoStop}%
\bibitem [{\citenamefont {Fisch}\ and\ \citenamefont
  {Boozer}(1980)}]{fisch_1980}%
  \BibitemOpen
  \bibfield  {author} {\bibinfo {author} {\bibfnamefont {N.~J.}\ \bibnamefont
  {Fisch}}\ and\ \bibinfo {author} {\bibfnamefont {A.~H.}\ \bibnamefont
  {Boozer}},\ }\bibfield  {title} {\enquote {\bibinfo {title} {Creating an
  asymmetric plasma resistivity with waves},}\ }\href {\doibase
  10.1103/PhysRevLett.45.720} {\bibfield  {journal} {\bibinfo  {journal} {Phys.
  Rev. Lett.}\ }\textbf {\bibinfo {volume} {45}},\ \bibinfo {pages} {720--722}
  (\bibinfo {year} {1980})}\BibitemShut {NoStop}%
\bibitem [{\citenamefont {Karney}, \citenamefont {Fisch},\ and\ \citenamefont
  {Jobes}(1985)}]{kfj}%
  \BibitemOpen
  \bibfield  {author} {\bibinfo {author} {\bibfnamefont {C.~F.~F.}\
  \bibnamefont {Karney}}, \bibinfo {author} {\bibfnamefont {N.~J.}\
  \bibnamefont {Fisch}}, \ and\ \bibinfo {author} {\bibfnamefont {F.~C.}\
  \bibnamefont {Jobes}},\ }\bibfield  {title} {\enquote {\bibinfo {title}
  {Comparison of the theory and the practice of lower-hybrid current drive},}\
  }\href {\doibase 10.1103/PhysRevA.32.2554} {\bibfield  {journal} {\bibinfo
  {journal} {Phys. Rev. A}\ }\textbf {\bibinfo {volume} {32}},\ \bibinfo
  {pages} {2554--2556} (\bibinfo {year} {1985})}\BibitemShut {NoStop}%
\bibitem [{\citenamefont {Maraschek}(2012)}]{Maraschek_2012}%
  \BibitemOpen
  \bibfield  {author} {\bibinfo {author} {\bibfnamefont {M.}~\bibnamefont
  {Maraschek}},\ }\bibfield  {title} {\enquote {\bibinfo {title} {Control of
  neoclassical tearing modes},}\ }\href {\doibase
  10.1088/0029-5515/52/7/074007} {\bibfield  {journal} {\bibinfo  {journal}
  {Nuclear Fusion}\ }\textbf {\bibinfo {volume} {52}},\ \bibinfo {pages}
  {074007} (\bibinfo {year} {2012})}\BibitemShut {NoStop}%
\bibitem [{\citenamefont {White}, \citenamefont {Gates},\ and\ \citenamefont
  {Brennan}(2015)}]{white_2015}%
  \BibitemOpen
  \bibfield  {author} {\bibinfo {author} {\bibfnamefont {R.~B.}\ \bibnamefont
  {White}}, \bibinfo {author} {\bibfnamefont {D.~A.}\ \bibnamefont {Gates}}, \
  and\ \bibinfo {author} {\bibfnamefont {D.~P.}\ \bibnamefont {Brennan}},\
  }\bibfield  {title} {\enquote {\bibinfo {title} {Thermal island
  destabilization and the greenwald limit},}\ }\href {\doibase
  10.1063/1.4913433} {\bibfield  {journal} {\bibinfo  {journal} {Physics of
  Plasmas}\ }\textbf {\bibinfo {volume} {22}},\ \bibinfo {pages} {022514}
  (\bibinfo {year} {2015})},\ \Eprint
  {http://arxiv.org/abs/https://doi.org/10.1063/1.4913433}
  {https://doi.org/10.1063/1.4913433} \BibitemShut {NoStop}%
\bibitem [{\citenamefont {Reiman}\ \emph {et~al.}(2020)\citenamefont {Reiman},
  \citenamefont {Bertelli}, \citenamefont {Bonoli}, \citenamefont {Fisch},
  \citenamefont {Frank}, \citenamefont {Jin}, \citenamefont {Nies},\ and\
  \citenamefont {Rodriguez}}]{allan}%
  \BibitemOpen
  \bibfield  {author} {\bibinfo {author} {\bibfnamefont {A.~H.}\ \bibnamefont
  {Reiman}}, \bibinfo {author} {\bibfnamefont {N.}~\bibnamefont {Bertelli}},
  \bibinfo {author} {\bibfnamefont {P.~T.}\ \bibnamefont {Bonoli}}, \bibinfo
  {author} {\bibfnamefont {N.~J.}\ \bibnamefont {Fisch}}, \bibinfo {author}
  {\bibfnamefont {S.~J.}\ \bibnamefont {Frank}}, \bibinfo {author}
  {\bibfnamefont {S.}~\bibnamefont {Jin}}, \bibinfo {author} {\bibfnamefont
  {R.}~\bibnamefont {Nies}}, \ and\ \bibinfo {author} {\bibfnamefont
  {E.}~\bibnamefont {Rodriguez}},\ }\href@noop {} {\enquote {\bibinfo {title}
  {Disruption avoidance via rf current condensation in magnetic islands
  produced by off-normal events},}\ } (\bibinfo {year} {2020}),\ \Eprint
  {http://arxiv.org/abs/2012.15389} {arXiv:2012.15389 [physics.plasm-ph]}
  \BibitemShut {NoStop}%
\bibitem [{\citenamefont {Lazzari}\ and\ \citenamefont
  {Westerhof}(2009)}]{De_Lazzari_2009}%
  \BibitemOpen
  \bibfield  {author} {\bibinfo {author} {\bibfnamefont {D.~D.}\ \bibnamefont
  {Lazzari}}\ and\ \bibinfo {author} {\bibfnamefont {E.}~\bibnamefont
  {Westerhof}},\ }\bibfield  {title} {\enquote {\bibinfo {title} {On the merits
  of heating and current drive for tearing mode stabilization},}\ }\href
  {\doibase 10.1088/0029-5515/49/7/075002} {\bibfield  {journal} {\bibinfo
  {journal} {Nuclear Fusion}\ }\textbf {\bibinfo {volume} {49}},\ \bibinfo
  {pages} {075002} (\bibinfo {year} {2009})}\BibitemShut {NoStop}%
\bibitem [{\citenamefont {Bertelli}, \citenamefont {Lazzari},\ and\
  \citenamefont {Westerhof}(2011)}]{Bertelli_2011}%
  \BibitemOpen
  \bibfield  {author} {\bibinfo {author} {\bibfnamefont {N.}~\bibnamefont
  {Bertelli}}, \bibinfo {author} {\bibfnamefont {D.~D.}\ \bibnamefont
  {Lazzari}}, \ and\ \bibinfo {author} {\bibfnamefont {E.}~\bibnamefont
  {Westerhof}},\ }\bibfield  {title} {\enquote {\bibinfo {title} {Requirements
  on localized current drive for the suppression of neoclassical tearing
  modes},}\ }\href {\doibase 10.1088/0029-5515/51/10/103007} {\bibfield
  {journal} {\bibinfo  {journal} {Nuclear Fusion}\ }\textbf {\bibinfo {volume}
  {51}},\ \bibinfo {pages} {103007} (\bibinfo {year} {2011})}\BibitemShut
  {NoStop}%
\bibitem [{\citenamefont {Fisch}(1987)}]{fisch_1987}%
  \BibitemOpen
  \bibfield  {author} {\bibinfo {author} {\bibfnamefont {N.~J.}\ \bibnamefont
  {Fisch}},\ }\bibfield  {title} {\enquote {\bibinfo {title} {Theory of current
  drive in plasmas},}\ }\href {\doibase 10.1103/RevModPhys.59.175} {\bibfield
  {journal} {\bibinfo  {journal} {Rev. Mod. Phys.}\ }\textbf {\bibinfo {volume}
  {59}},\ \bibinfo {pages} {175--234} (\bibinfo {year} {1987})}\BibitemShut
  {NoStop}%
\bibitem [{\citenamefont {Erba}\ \emph {et~al.}(1998)\citenamefont {Erba},
  \citenamefont {Aniel}, \citenamefont {Basiuk}, \citenamefont {Becoulet},\
  and\ \citenamefont {Litaudon}}]{Erba_1998}%
  \BibitemOpen
  \bibfield  {author} {\bibinfo {author} {\bibfnamefont {M.}~\bibnamefont
  {Erba}}, \bibinfo {author} {\bibfnamefont {T.}~\bibnamefont {Aniel}},
  \bibinfo {author} {\bibfnamefont {V.}~\bibnamefont {Basiuk}}, \bibinfo
  {author} {\bibfnamefont {A.}~\bibnamefont {Becoulet}}, \ and\ \bibinfo
  {author} {\bibfnamefont {X.}~\bibnamefont {Litaudon}},\ }\bibfield  {title}
  {\enquote {\bibinfo {title} {Validation of a new mixed bohm/gyro-bohm model
  for electron and ion heat transport against the {ITER}, tore supra and
  {START} database discharges},}\ }\href {\doibase 10.1088/0029-5515/38/7/305}
  {\bibfield  {journal} {\bibinfo  {journal} {Nuclear Fusion}\ }\textbf
  {\bibinfo {volume} {38}},\ \bibinfo {pages} {1013--1028} (\bibinfo {year}
  {1998})}\BibitemShut {NoStop}%
\bibitem [{\citenamefont {Zohm}\ \emph {et~al.}(2007)\citenamefont {Zohm},
  \citenamefont {Gantenbein}, \citenamefont {Leuterer}, \citenamefont {Manini},
  \citenamefont {Maraschek}, \citenamefont {Yu},\ and\ \citenamefont {the ASDEX
  Upgrade~Team}}]{Zohm_2007}%
  \BibitemOpen
  \bibfield  {author} {\bibinfo {author} {\bibfnamefont {H.}~\bibnamefont
  {Zohm}}, \bibinfo {author} {\bibfnamefont {G.}~\bibnamefont {Gantenbein}},
  \bibinfo {author} {\bibfnamefont {F.}~\bibnamefont {Leuterer}}, \bibinfo
  {author} {\bibfnamefont {A.}~\bibnamefont {Manini}}, \bibinfo {author}
  {\bibfnamefont {M.}~\bibnamefont {Maraschek}}, \bibinfo {author}
  {\bibfnamefont {Q.}~\bibnamefont {Yu}}, \ and\ \bibinfo {author}
  {\bibnamefont {the ASDEX Upgrade~Team}},\ }\bibfield  {title} {\enquote
  {\bibinfo {title} {Control of {MHD} instabilities by {ECCD}: {ASDEX} upgrade
  results and implications for {ITER}},}\ }\href {\doibase
  10.1088/0029-5515/47/3/010} {\bibfield  {journal} {\bibinfo  {journal}
  {Nuclear Fusion}\ }\textbf {\bibinfo {volume} {47}},\ \bibinfo {pages}
  {228--232} (\bibinfo {year} {2007})}\BibitemShut {NoStop}%
\bibitem [{\citenamefont {Volpe}\ \emph {et~al.}(2009)\citenamefont {Volpe},
  \citenamefont {Austin}, \citenamefont {La~Haye}, \citenamefont {Lohr},
  \citenamefont {Prater}, \citenamefont {Strait},\ and\ \citenamefont
  {Welander}}]{Volpe_2009}%
  \BibitemOpen
  \bibfield  {author} {\bibinfo {author} {\bibfnamefont {F.~A.~G.}\
  \bibnamefont {Volpe}}, \bibinfo {author} {\bibfnamefont {M.~E.}\ \bibnamefont
  {Austin}}, \bibinfo {author} {\bibfnamefont {R.~J.}\ \bibnamefont {La~Haye}},
  \bibinfo {author} {\bibfnamefont {J.}~\bibnamefont {Lohr}}, \bibinfo {author}
  {\bibfnamefont {R.}~\bibnamefont {Prater}}, \bibinfo {author} {\bibfnamefont
  {E.~J.}\ \bibnamefont {Strait}}, \ and\ \bibinfo {author} {\bibfnamefont
  {A.~S.}\ \bibnamefont {Welander}},\ }\bibfield  {title} {\enquote {\bibinfo
  {title} {Advanced techniques for neoclassical tearing mode control in
  {DIII-D}},}\ }\href {\doibase 10.1063/1.3232325} {\bibfield  {journal}
  {\bibinfo  {journal} {Physics of Plasmas}\ }\textbf {\bibinfo {volume}
  {16}},\ \bibinfo {pages} {102502} (\bibinfo {year} {2009})}\BibitemShut
  {NoStop}%
\bibitem [{\citenamefont {Nelson}\ \emph {et~al.}(2019)\citenamefont {Nelson},
  \citenamefont {Haye}, \citenamefont {Austin}, \citenamefont {Welander},\ and\
  \citenamefont {Kolemen}}]{Nelson_2019}%
  \BibitemOpen
  \bibfield  {author} {\bibinfo {author} {\bibfnamefont {A.}~\bibnamefont
  {Nelson}}, \bibinfo {author} {\bibfnamefont {R.~L.}\ \bibnamefont {Haye}},
  \bibinfo {author} {\bibfnamefont {M.}~\bibnamefont {Austin}}, \bibinfo
  {author} {\bibfnamefont {A.}~\bibnamefont {Welander}}, \ and\ \bibinfo
  {author} {\bibfnamefont {E.}~\bibnamefont {Kolemen}},\ }\bibfield  {title}
  {\enquote {\bibinfo {title} {Simultaneous detection of neoclassical tearing
  mode and electron cyclotron current drive locations using electron cyclotron
  emission in {DIII-D}},}\ }\href {\doibase
  https://doi.org/10.1016/j.fusengdes.2019.02.089} {\bibfield  {journal}
  {\bibinfo  {journal} {Fusion Engineering and Design}\ }\textbf {\bibinfo
  {volume} {141}},\ \bibinfo {pages} {25 -- 29} (\bibinfo {year}
  {2019})}\BibitemShut {NoStop}%
\bibitem [{\citenamefont {Frank}\ \emph {et~al.}(2020)\citenamefont {Frank},
  \citenamefont {Reiman}, \citenamefont {Fisch},\ and\ \citenamefont
  {Bonoli}}]{Frank_2020}%
  \BibitemOpen
  \bibfield  {author} {\bibinfo {author} {\bibfnamefont {S.}~\bibnamefont
  {Frank}}, \bibinfo {author} {\bibfnamefont {A.}~\bibnamefont {Reiman}},
  \bibinfo {author} {\bibfnamefont {N.}~\bibnamefont {Fisch}}, \ and\ \bibinfo
  {author} {\bibfnamefont {P.}~\bibnamefont {Bonoli}},\ }\bibfield  {title}
  {\enquote {\bibinfo {title} {Generation of localized lower-hybrid current
  drive by temperature perturbations},}\ }\href {\doibase
  10.1088/1741-4326/aba3fc} {\bibfield  {journal} {\bibinfo  {journal} {Nuclear
  Fusion}\ }\textbf {\bibinfo {volume} {60}},\ \bibinfo {pages} {096027}
  (\bibinfo {year} {2020})}\BibitemShut {NoStop}%
\bibitem [{\citenamefont {Westerhof}, \citenamefont {de~Blank},\ and\
  \citenamefont {Pratt}(2016)}]{Westerhof_2016}%
  \BibitemOpen
  \bibfield  {author} {\bibinfo {author} {\bibfnamefont {E.}~\bibnamefont
  {Westerhof}}, \bibinfo {author} {\bibfnamefont {H.}~\bibnamefont {de~Blank}},
  \ and\ \bibinfo {author} {\bibfnamefont {J.}~\bibnamefont {Pratt}},\
  }\bibfield  {title} {\enquote {\bibinfo {title} {New insights into the
  generalized rutherford equation for nonlinear neoclassical tearing mode
  growth from 2{D} reduced {MHD} simulations},}\ }\href {\doibase
  10.1088/0029-5515/56/3/036016} {\bibfield  {journal} {\bibinfo  {journal}
  {Nuclear Fusion}\ }\textbf {\bibinfo {volume} {56}},\ \bibinfo {pages}
  {036016} (\bibinfo {year} {2016})}\BibitemShut {NoStop}%
\bibitem [{\citenamefont {Sauter}(2004)}]{Sauter_2004}%
  \BibitemOpen
  \bibfield  {author} {\bibinfo {author} {\bibfnamefont {O.}~\bibnamefont
  {Sauter}},\ }\bibfield  {title} {\enquote {\bibinfo {title} {On the
  contribution of local current density to neoclassical tearing mode
  stabilization},}\ }\href {\doibase 10.1063/1.1787791} {\bibfield  {journal}
  {\bibinfo  {journal} {Physics of Plasmas}\ }\textbf {\bibinfo {volume}
  {11}},\ \bibinfo {pages} {4808--4813} (\bibinfo {year} {2004})}\BibitemShut
  {NoStop}%
\end{thebibliography}%

\end{document}